\documentclass[reprint,superscriptaddress,amsmath,amssymb,showkeys,aps,prl]{revtex4-2}
\usepackage{upgreek}
\usepackage{graphicx}
\usepackage{dcolumn}
\usepackage{bm}

\usepackage{academicons}

\usepackage{float}
\begin{document}

\title{Tuning the domain wall conductivity in bulk lithium niobate by uniaxial stress}

\author{Ekta Singh}
\affiliation{Institute of Applied Physics, TU Dresden, N{\"o}thnitzer Stra{\ss}e 61, 01187 Dresden, Germany }
\affiliation{Max Planck Institute for Chemical Physics of Solids, N{\"o}thnitzer Stra{\ss}e 40, 01187 Dresden, Germany}
\author{Henrik Beccard}

\affiliation{Institute of Applied Physics, TU Dresden, N{\"o}thnitzer Stra{\ss}e 61, 01187 Dresden, Germany }

\author{Zeeshan H. Amber}

\affiliation{Institute of Applied Physics, TU Dresden, N{\"o}thnitzer Stra{\ss}e 61, 01187 Dresden, Germany }

\author{Julius Ratzenberger}

\affiliation{Institute of Applied Physics, TU Dresden, N{\"o}thnitzer Stra{\ss}e 61, 01187 Dresden, Germany }

\author{Clifford W. Hicks}
\affiliation{Max Planck Institute for Chemical Physics of Solids, N{\"o}thnitzer Stra{\ss}e 40, 01187 Dresden, Germany} 
\affiliation{School of Physics and Astronomy, University of Birmingham, Birmingham B15 2TT, United Kingdom}
\author{Michael R{\"u}sing}

\affiliation{Institute of Applied Physics, TU Dresden, N{\"o}thnitzer Stra{\ss}e 61, 01187 Dresden, Germany }

\author{Lukas M. Eng}

\affiliation{Institute of Applied Physics, TU Dresden, N{\"o}thnitzer Stra{\ss}e 61, 01187 Dresden, Germany }
 
\affiliation{ct.qmat: Dresden-W{\"u}rzburg Cluster of Excellence-EXC 2147, TU Dresden, 01062 Dresden, Germany}

\date{\today}

\begin{abstract}
Conductive domain walls (CDWs) in insulating ferroelectrics have recently attracted considerable attention due to their unique topological, optical, and electronic properties, and offer potential applications such as in memory devices or re-writable circuitry. The electronic properties of domain walls (DWs) can be tuned by the application of strain, hence controlling the charge carrier density at DWs. In this work, we study the influence of uniaxial stress on the conductivity of DWs in the bulk single crystal lithium niobate (LiNbO$_3$). Using conductive atomic force microscopy (cAFM), we observe a large asymmetry in the conductivity of DWs, where only negatively screened walls, so called head-to-head DWs, are becoming increasingly conductive, while positively screened, tail-to-tails DWs, show a decrease in conductivity. This asymmetry of DW conductivity agrees with our theoretical model based on the piezoelectric effect. In addition, we observed that the current in the DW increases up to an order of magnitude for smaller compressive stresses of 100 MPa. This response of DWs remained intact for multiple stress cycles over 2 months, opening a path for future applications.\end{abstract}

\keywords{Ferroelectric conductive domain walls (CDWs), lithium niobate (LiNbO$_3$), uniaxial stress, conductive atomic force microscopy (cAFM), piezoresponse force microscopy (PFM)}

\maketitle
\section{Introduction}\label{sec1}
\begin{figure*}
\centering
\includegraphics{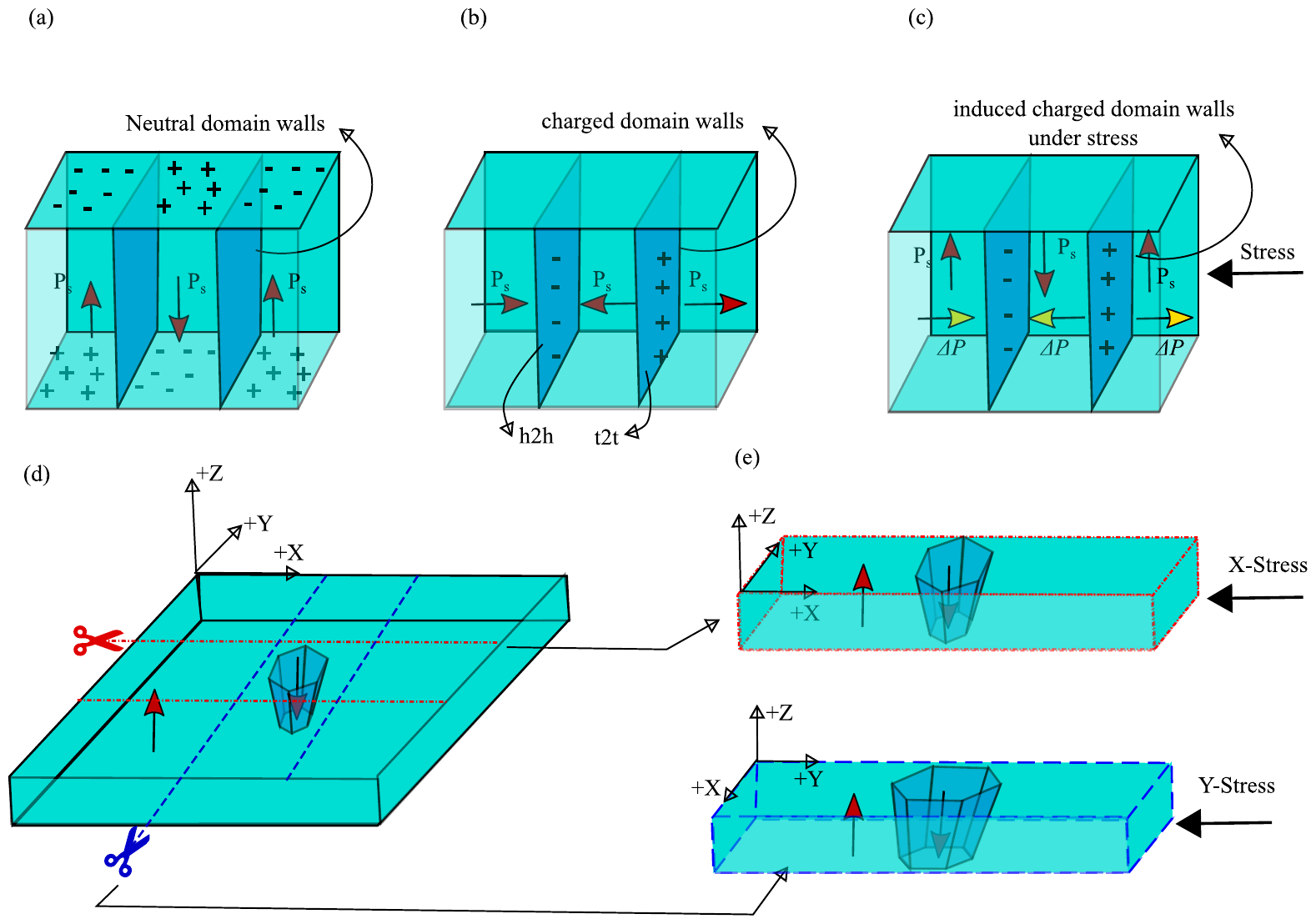}
\caption{\label{fig:domain_concept}Sketch: (a) Neutral domain walls in a ferroelectric, where red arrows are representing the spontaneous polarization parallel to domain walls. (b) Charged domain walls: spontaneous polarizations meet head-to-head (h2h) and tail-to-tail (t2t) perpendicular to domain walls. (c) Neutral domain walls but charged due to induced polarization, represented by yellow arrows, when stress is applied to a crystal. Charges shown in the images are screening charge carriers. (d) Piece of single crystal Z-cut LiNbO$_3$ containing a hexagonal domain. (e) Two different LiNbO$_3$ samples cut along different axes from the parent crystal in image (d). The samples are cut such that stress can be applied to the crystallographic x and y axes.}
\end{figure*}

For the last decade, ferroelectric domain walls (DWs) have been in the focus of research due to their outstanding optical, electrical, and topological properties, that promise numerous applications such as resistive switches and non-volatile ferroelectric memory devices \cite{Kirbus2019,Sharma2017a,Zhao2020}. These applications take advantage of conductive nature of ferroelectric DWs, where on and off states of the devices can be defined by the resistance of DWs \cite{Sharma2017a,Kampfe2020b,Jiang2018,Li2016,Gruverman2008,Catalan2012}. In most models, the conductivity of DWs is connected to the order parameter of the surrounding domains, which for ferroelectrics is the spontaneous polarization $P_s$. In the context of conductivity, three main configurations of a DW are distinguished: 

(1) Neutral DWs; when $P_s$ from neighbouring domains are aligned antiparallel to each other as depicted in Fig.~\ref{fig:domain_concept}(a).

(2) head-to-head (h2h) DWs; when $P_s$ from different domains meet at their positive ends; and 

(3) tail-to-tail (t2t) DWs; which are the opposing configuration of h2h, see Fig.~\ref{fig:domain_concept}(b).  

The convergence of polarization in cases (2) and (3), creates non-zero bound surface charges localized at the DW, which becomes a source of the so-called depolarization field. This field is then compensated by mobile screening  charge carriers such as electrons, holes, polarons, or mobile ions. 
 In some cases the depolarization field is even strong enough ($\approx$ 1 MV/cm) to locally bend the conduction band  below the Fermi level, hence creating a 2D electron gas at the DW, which, for example, was reported in BaTiO$_3$ \cite{SeidelEditora, Sluka2012}. The presence of such charged CDWs is reported for many ferroelectrics such as BiFeO$_3$ \cite{Condurache2021,Rojac2017}, PbTiO$_3$ \cite{Stolichnov2015}, BaTiO$_3$ \cite{Sluka2013}, HoMnO$_3$ \cite{Wu2012}, LiNbO$_3$ \cite{Gonnissen2016,Werner2017,Schroder2012} etc.

In ferroelectrics, DWs can be easily written, erased, moved, or even switched between different states of conductivity. 
The most common method is by the application of electric fields.
Electric fields are able to create or erase DWs via ferroelectric poling. In case of LiNbO$_3$ it further allows to control the amount of charge accumulation, e.g. by controlling the tilt angle of the DW with respect to the polar axis \cite{Godau2017,Kampfe2020b,Lu2019}. The control of charge accumulation, or in other words, the conductivity of DWs allows to enlarge the memory window for ferroelectric based memory devices \cite{Kampfe2020b}. Another elegant possibility to control and study the accumulation of bound charges is via direct piezoelectricity \cite{Chen2012,Alsubaie2017}, i.e. by inducing an additional polarization component through stress or strain [see Fig.~\ref{fig:domain_concept}(c)], which so far has only been reported for a few selected materials. For example, Ederer et al. evaluated the effect of epitaxial strain of up to $\pm 2.5\%$ on the spontaneous polarization ($P_s$) for different ferroelectrics \cite{Ederer2005}. Experimentally, Chen et. al. measured the spontaneous polarization and conductivity of DWs of strained BiFO$_3$ thin films. They reported that strain tuning changes the DW conductivity by several orders of magnitude indicating effects not just due to screening charges but band-bending as well \cite{Chen2018}. In the reported experimental or theoretical cases, the application and amount of strain is limited by lattice-mismatched epitaxial growth of thin films on specifically selected substrates and, therefore, cannot be generalized to all ferroelectric materials or crystallographic orientations \cite{Schlom2007}. 

To enable the study of strain on the DW conductivity for any bulk ferroelectric, we report the room temperature domain wall conductivity of LiNbO$_3$ under uniaxial stress, by combining a in-situ tunable uniaxial stress cell with scanning probe microscopy. With the help of conductive atomic force microscopy (cAFM), we show the local change of the current distribution in DWs, when stress is applied along different crystallographic directions in LiNbO$_3$. The experimental results are readily explained with our model based on the direct piezoelectric effect in LiNbO$_3$.

For our study, we have chosen the ferroelectric material lithium niobate (5 $\%$ MgO-doped LiNbO$_3$), where highly CDWs with currents of up to $\approx$1 mA at 10 V in 200 $\upmu$m thick crystals have been reported \cite{Godau2017, Kirbus2019a}. Recent experiments in LNO have demonstrated that DWs can be switched between conductive and non-conductive states with a memory window of $>10^4$ via electric fields in both, thin films and bulk devices \cite{Kirbus2019a,Kampfe2020b}. This switching process can be used for fabricating a two-terminal memory device with an extrapolated 80\%-lifetime of $>10$ years \cite{Kampfe2020b}. Therefore, LiNbO$_3$ is an ideal model system to study the effects of strain on DW conductivity.

\section{Results}\label{sec2}
\begin{figure*}
\centering
\includegraphics[width=7in]{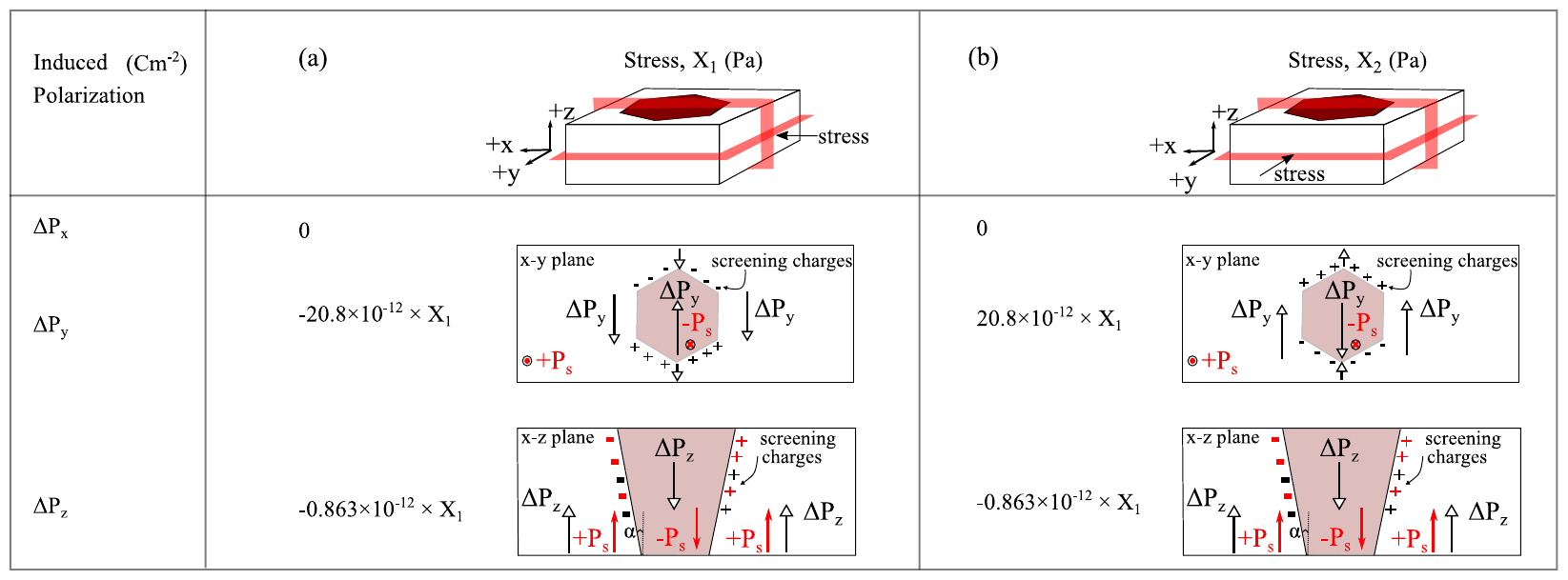}
\caption{\label{fig:model_sketch} Projection of induced polarization ($P_j$ = $d_{jkl}$$X_{kl}$) at DWs along different axes at zero external electric field when LiNbO$_3$ is stressed along (a) x-axis, and (b) y-axis. Where for LiNbO$_3$, $d_{22}$ = 20.8 $\times$ 10$^{-12}$ C/N and $d_{31}$ = -0.863 $\times$ 10$^{-12}$ C/N at room temperature \protect\cite{Yamada1967}. The sign of stress should be negative for compression and positive for tension. Each sketch belongs to the respective column above depicts directions of induced polarization at DWs for compression configuration. The directions should be reversed for tension. Charges shown at DWs are screening charges, which are responsible for the conductivity in DWs (detailed calculations are given in supplemantary)}
\end{figure*}
The samples used in this measurement are single-crystal Z-cut 5\% MgO-doped LiNbO$_3$ (LNO). For this study we analyzed two samples, which are prepared such that stress can be applied along different crystal axes [see Fig.~\ref{fig:domain_concept}(d) and (e)]. Sample LNO-01 is compressed along the crystallographic x-axis, while sample LNO-02 is compressed along the crystallographic y-axis. Both these samples carry a conductive hexagonally shaped DW, which was first rendered more conductive (enhanced DWs) using the enhancement procedure  by Godau et.al. \cite{Godau2017}(see method section). The samples were stressed with the help of a uniaxial stress cell, which is based on piezoelectric stacks \cite{Hicks2014}. This cell can apply controlled, user defined tensile or compressive stress in a continuous fashion on the bulk samples (see method section). Application of the stress along these two directions results in principally different behaviors of the DW conductivity of LiNbO$_3$, as explained in section \ref{sec:model} by the theoretical model. 

\subsection{Theoretical model} \label{sec:model}
\noindent The conductivity of ferroelectric DWs in LiNbO$_3$ is believed to be related to the amount of screening charges present at the DW. As shown in Fig. S2 and S3 of supplement subsection S1.1 and discussed in the literature \cite{Godau2017,Schroder2012,Kirbus2019a} any increase of the DW inclination angle $\alpha$, with respect to the polar axis results in an increase of the screening charge carrier density $\sigma$ and, ideally its conductivity by $\sigma = 2P_{s}\sin{\alpha}$. Alike, applying stress to the sample results in a change of screening charge carriers by induced polarization through the direct piezoelectric effect. This can be used to predict and describe the behavior of CDWs with respect to applied stress. The polarization $\Delta P$ in the crystal of LiNbO$_3$ can be described via:
\begin{equation}{\label{eq:tensor}}
	\begin{bmatrix}
		P_x\\P_y \\P_z 
	\end{bmatrix}
	= 
	\begin{bmatrix}
		0&0&0&0&d_{15}&-2d_{22}\\-d_{22}&d_{22}&0&d_{15}&0&0\\d_{31}&d_{31}&d_{33}&0&0&0
	\end{bmatrix}
	\begin{bmatrix}
		X_1\\X_2\\X_3\\X_4\\X_5\\X_6
	\end{bmatrix}
\end{equation}
\begin{figure*}
\includegraphics{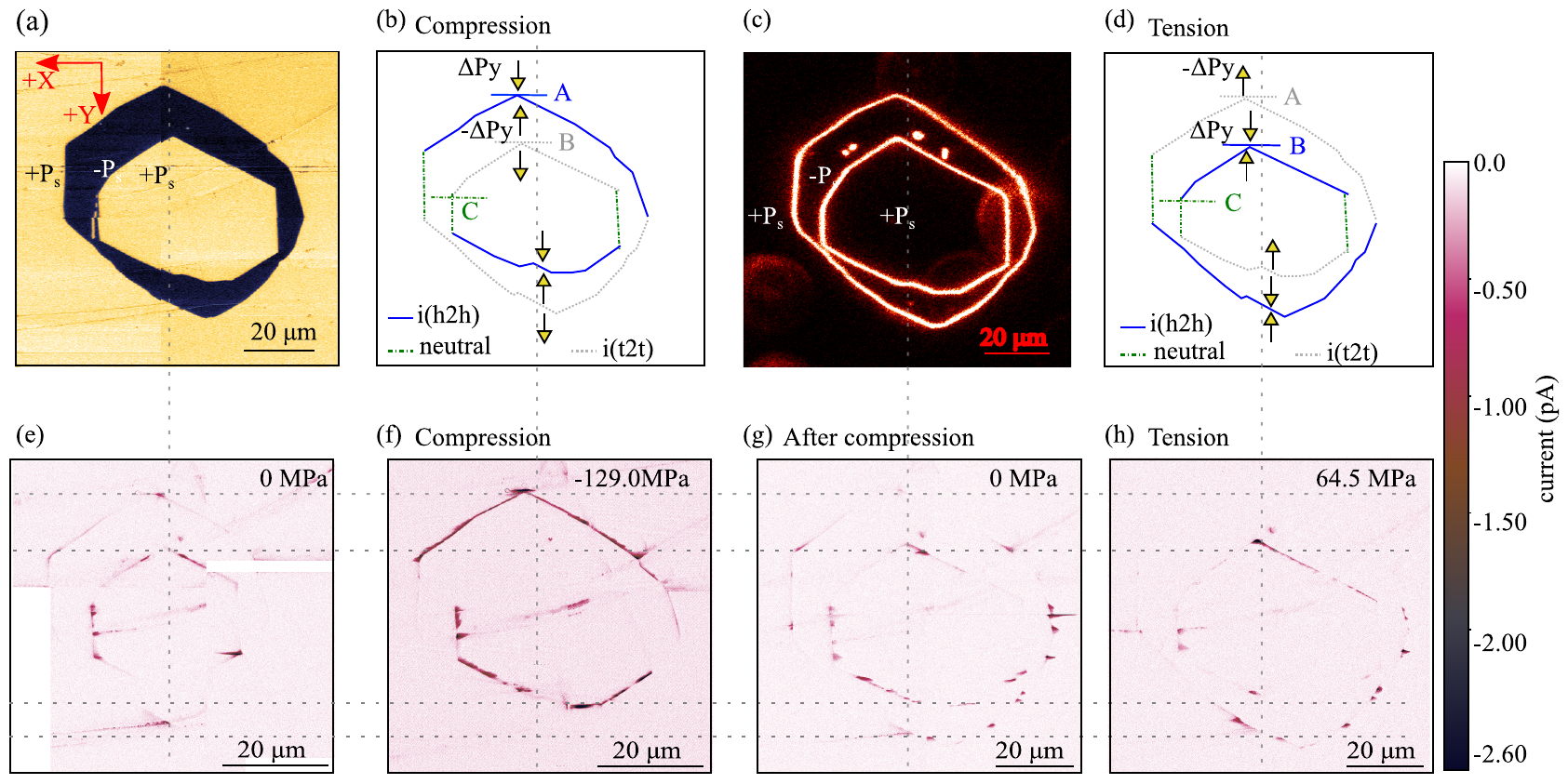}
\caption{\label{fig:domain wall}(a) Stitched PFM phase image of LNO-01 sample (x-compression), arrows on top left shows crystallographic axes (b) Current distribution in DWs under compressive stress according to a model based on direct piezoelectricity (c) SHG image of LNO sample. (d) Current distribution of DWs under tensile stress according to a model based on direct piezoelectricity; Stitched cAFM image of DWs: (e) at 0 MPa stress, (f) at -129.0 MPa compressive stress, (g) at 0 MPa after compression, and (h) at 64.5 MPa tensile stress. The DWs in (f) and (h) shows expected response as in sketches above them (b) and (d) respectively.}
\end{figure*}

\noindent where $[d]$ is the matrix of piezoelectric strain coefficients and $[X]$ is the stress matrix \cite{Yamada1967,Smith1971a,Weis1985} (details on the calculation and the tensor elements for LiNbO$_3$ are presented in the supplement section S2). When the LiNbO$_3$ crystal is stressed along the crystallographic x-axis as in the sample LNO-01 which we have measured in this work, an extra polarization is induced along both the y- and z-axes, as provided in Fig~\ref{fig:model_sketch}(a). The values of induced polarizations along the y-axis ($\Delta P_y = 0.208 \times 10^{-2}$ Cm$^{-2}$) and the z-axis ($\Delta P_z = 0.863 \times 10^{-4}$ Cm$^{-2}$) for a compressive stress of -100 MPa are significantly smaller than the spontaneous polarization of LiNbO$_3$ ($P_s$ $\approx$ 0.7 Cm$^{-2}$ \cite{Wemple1968a}). However, as the polarization $\Delta P_y$ is projected perpendicular to $\Delta P_z$ at the DW, it contributes as a cosine component to the surface charge density equation and results in a significant surface charge density ($\sigma$) at the DWs ( $\Delta \sigma_y + \Delta \sigma_z $ = 2$\cdot$$\Delta$P$_{y} \cos {\alpha}$ +2$\cdot$$\Delta$P$_{z} \sin {\alpha}$), as depicted by the sketches in Fig.~\ref{fig:model_sketch}(a). This is in the same order of magnitude as the natural charge density $\sigma$  for small angles of inclination of $\alpha = 1^{\circ}$, typically observed for enhanced domain walls \cite{Godau2017}. As a result, x-compressed DWs, which are oriented at some angle to the y-axis, should become h2h or t2t like, and thus should be screened additionally  by negative or positive mobile charge carriers, respectively. Depending on their geometry, from now on we will refer to them as induced h2h, i(h2h), and induced t2t, i(t2t) DWs. 

Since LiNbO$_3$ is piezoelectric along the y-axis, one should expect exactly the opposite behaviour when the sample is compressed along that crystallographic y-axis, as we will indeed report below for sample LNO-02. This happens due to the sign inversion of $d_{22}$ in equation~\ref{eq:tensor}. This is depicted in Fig.~\ref{fig:model_sketch}(b). In a y-compression scenario, the induced polarization $\Delta P_y$ will change the direction by 180$^{\circ}$ in contrast to the $\Delta P_y$ in x-compression, and the DWs which were i(h2h) in x-compression will become i(t2t) for y-compression.

When applying this model to our LNO-01 and LNO-02 samples, we expect DW currents to behave as described in the sketches of Fig.~\ref{fig:domain wall} (b), (d) and Fig.~\ref{fig:LNO2}(b). We show that the DWs highlighted in blue-solid lines should become i(h2h) type (negatively screened) while grey dotted DWs should become i(t2t) type (positively screened), and green dashed DWs should not be influenced by uniaxial stress at all. The opposite behaviour should be observed under tensile stress.

\subsection{Experiment: domain walls under uniaxial stress}

In order to locate the domains, piezo response force microscopy (PFM) was performed on the +z side of sample LNO-01, as illustrated in Fig.~\ref{fig:domain wall} (a). As seen in Fig.~\ref{fig:domain wall} (c), the PFM image agree well with SHG imaging and confirm the presence of the same DWs. In the PFM image, yellow and black color contrast represents a phase difference of $180^{\circ}$ between two different orientations, while in SHG microscopy the presence of DWs is indicated by an enhanced backscattered signal. Based on the PFM scans, cAFM was performed at the same location at different stresses while applying a -10 V dc voltage to the bottom contact (-z side). Fig.~\ref{fig:domain wall}(e) shows a cAFM overview at 0 MPa. Here, only parts of the DWs show conductivity. The reason for this could be different near surface inclination angles of DWs, leading to locally different Schottky barriers, which has been reported earlier \cite{Godau2017,Wu2012,Lu2019,Schroder2012}. However, the observed locations correspond to the shape and location of the DWs as observed in PFM and SHG microscopy.

When compressing the sample along the x-axis, we expect the DWs with i(h2h) and i(t2t) configuration to be additionally charged. The i(h2h) DWs must get more conductive because the amount of negative screening charges increases. On the other hand i(t2t) DWs first should fully compensate the preexisting negative screening charges at the DW. This means for i(t2t) DWs that the current should first reduce to zero and on application of further compressive stress, one should expect these i(t2t) DWs to become conductive again. Both types of DWs are highlighted by blue-solid and grey-dotted lines in the sketch in Fig.~\ref{fig:domain wall}(b). The image in Fig.~\ref{fig:domain wall}(f) below shows a cAFM scan taken at -129 MPa. Indeed, we see an enhanced conductivity for the i(h2h) configurations, while the i(t2t) walls are observed to show a decreasing conductivity. When the compression is relaxed back to 0 MPa again, as depicted in Fig.~\ref{fig:domain wall}(g), it retains a qualitatively similar picture to the initial state [Fig.~\ref{fig:domain wall}(e)]. When tensile stress is applied, only the walls with a i(h2h) configuration with respect to the induced polarization exhibit a significant increased conductivity, while the induced i(t2t) DWs show a disappearing conductivity as shown in Fig.~\ref{fig:domain wall}(d) and (h). In all cases, the neutral walls with respect to the induced polarization show a similar qualitative and quantitative behavior, independent of the applied stress. Additional experiments were performed in intermediate steps of approximately 16.12 MPa for both, tension and compression in the range from -129 MPa (compression) up to +64.5 MPa (tensile), as well as repeated multiple times. Selected results can be found in Fig. S6 of the supplement subsection S3.1. In all cases, similar qualitative and quantitative results were obtained as expected, where only the i(h2h) DWs show a significant contribution to the conductivity. The observation that only h2h DWs exhibit a high electric conductivity has been reported before for LiNbO$_3$ and other ferroelectrics for tilted DWs, and can be explained by the proposed microscopic mechanism of DW conductivity. For LiNbO$_3$, the DW conductivity is explained by hoping transport of electrons in bound-polaronic states, while hole polarons are expected to be only a weak contributor. Hence, only h2h, i.e. negatively screened walls, will contribute to overall conductivity. In this regard, our experiment is in agreement with the polaron-hopping transport mechanism \cite{Xiao2018}.
\begin{figure}
\centering
\includegraphics{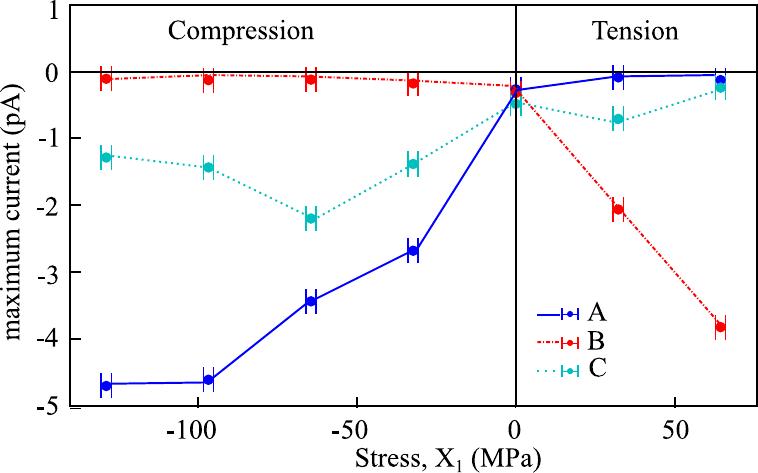}
\caption{\label{fig:profile}Change in current with stress along in line profiles A, B, and C taken from different sections of DWs of sample LNO-01(x-compression), provided in Fig.~\ref{fig:domain wall} (b) and (d)}
\end{figure}
Based on the piezoelectric theory the induced polarization is directly proportional to the applied stress. Hence, the conductivity for h2h walls should increase approximately linear with increasing polarization. In order to show the change in the current as function of stress more clearly, we have plotted in Fig. \ref{fig:profile} the maximum current from line profiles A, B and C (taken from i(h2h), i(t2t) and neutral parts of DWs in Fig. \ref{fig:domain wall}, respectively). This graph shows  three different kinds of behavior:

 (1) The DW depicted along the line profile A shows an increase in conductivity on compression, while it shows almost no response for tensile stress within the resolution limit of our setup. 
 
 (2) In contrast, the DW in profile B shows the opposite response, as this wall gets an induced i(h2h) configuration for tensile stress. On the other hand,  
 
 (3) The DW in line profile C shows no distinctive behavior. 
 
 It should be noted that in this experiment additional to the effects of induced polarizations $\Delta$P$_y$ and $\Delta$P$_z$, we   also expect to see effects from local roughness and local inclinations of the DWs. Therefore, charge carriers inside the crystal might not move in a simple path vertically along the DWs and therefore distorting the ideally expected linear relationships. Effects like this may be predicted based on a resistor network model of DWs \cite{Wolba2018} in future work. 
 
 Apart from the cAFM studies on the sample LNO-01, we have also investigated the response of the DWs to stress with large, deposited electrodes covering all the DWs. These electrodes were the same that were also used for the enhancement procedure. In this macroscopic picture, we see an overall increase in conductivity with applied stress similar to Fig. 4, line profile A for LNO-01. This is in agreement with the fact that in this specific sample the i(h2h) DWs dominate the conductivity. However, it should be noted that this result may not be generalized to all samples and will depend on which induced domain type dominates [i(h2h) or i(t2t)] the conductivity, as was observed for sample LNO-02(details  to macroscopic curve for both samples are provided in Fig. S2 and S3 of the supplement section S1.1). Therefore, this retrospectively motivates the microscopic cAFM study.

\begin{figure}
\includegraphics{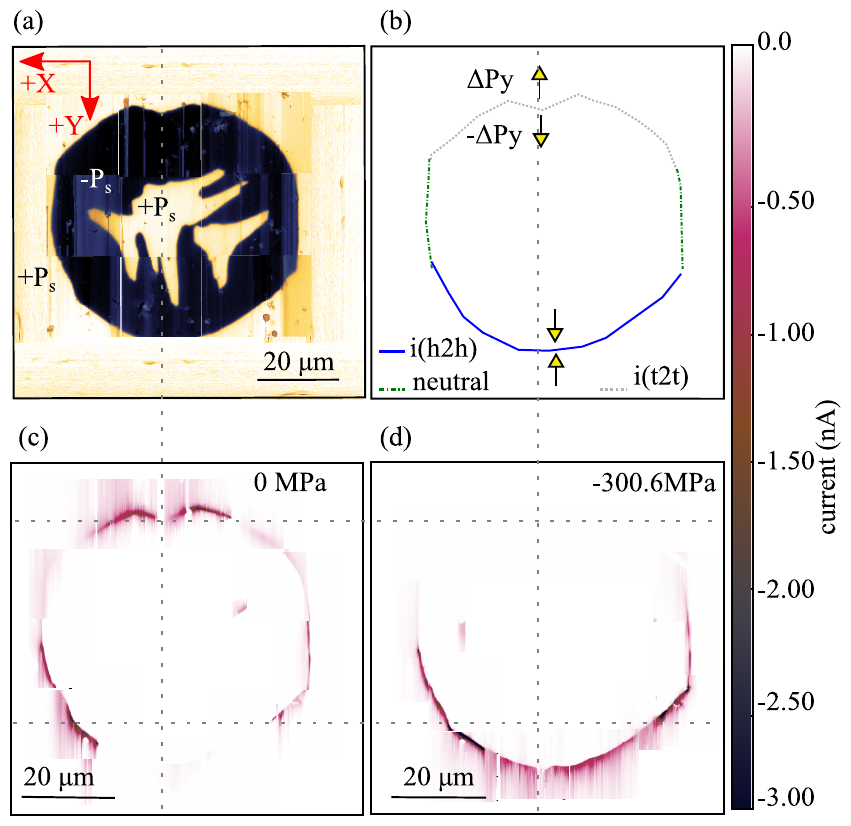}
\caption{\label{fig:LNO2}(a) Stitched PFM phase image of LNO-02 sample (y-compression), arrows on top left shows crystallographic axes (b) Current distribution in DWs under compressive stress according to a model based on direct piezoelectricity, predicts opposite response to the sample LNO-01. Stitched cAFM image of DWs: (c) at 0 MPa stress, (d) at -300.6 MPa compressive stress.}
\end{figure}

For a sample compressed along y-axis different behaviour to x-axis compression according to the model is expected. Hence, when the similar cAFM measurement was performed on sample LNO-02, we observed the opposite response as was explained by the theoretical model in last section. Fig.~\ref{fig:LNO2}(a) represents the PFM image taken on +z surface. The DWs in sample LNO-02 are more conductive by factor of 3 even at 0 MPa, as compared to sample LNO-01 as shown by Fig.~\ref{fig:LNO2}(c). This is due to the inclination angle $\alpha$ of DWs in LNO-02 is more than in LNO-01 by 1$^\circ$, see angle calculation in Fig S4 and S5 of supplement subsection S1.2. We only show the cAFM image of the outer wall in sample LNO-02 because of the irregular shape of the inner DW. Nevertheless, when compressive stress of -300.6 MPa was applied along the y-axis, the walls along the +y direction of the crystal becomes i(h2h) and hence more conductive which is opposite to the response of sample LNO-01 as shown by in Fig.~\ref{fig:LNO2}(d).

(Further information on sample LNO-02 can be found in the supplement).
\section{Methods}\label{sec11}
\subsection{Stress cell}
\begin{figure*}[t]
\includegraphics[width=160mm]{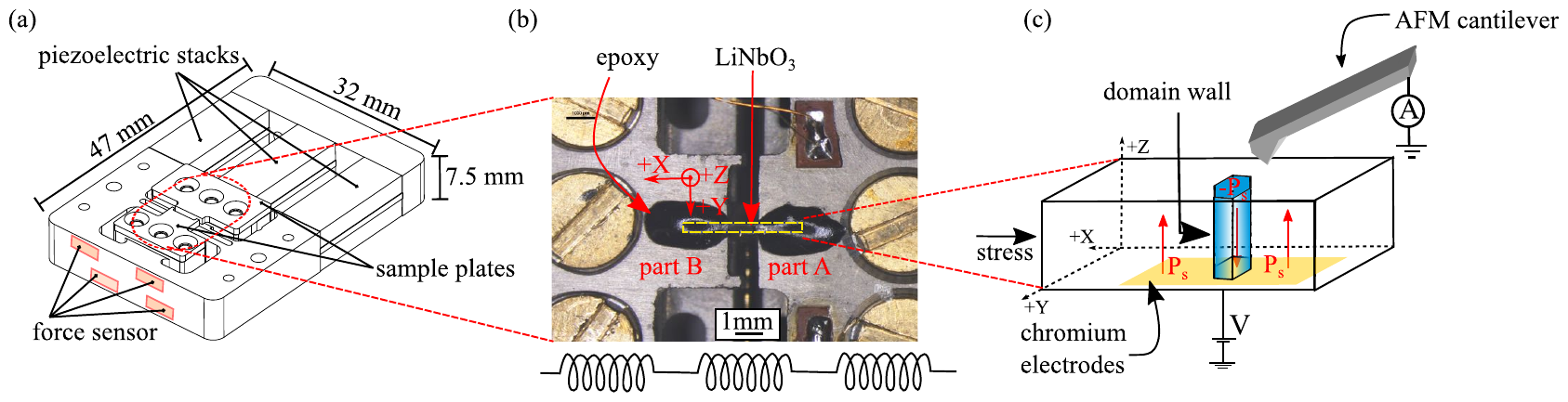}	
\caption{\label{fig:cell}(a) 3-dimensional sketch of the uniaxial stress cell. (b) Zoom in image of the stress cell: Z-cut 5\% MgO-doped LiNbO$_3$ (sample: LNO-01) mounted on the cell with 2850FT epoxy. (c) Sketch of the setup for cAFM measurement with applied stress. Domains and domain walls are electrically connected by a 20 nm thick chromium electrode at the -z side and grounded by the cantilever at the +z side.}
\end{figure*}

The stress cell used in this work has a central piezoelectric stack, connected to the main body of the cell, which applies  compressive stress, while the outer stacks apply tensile stress \cite{Hicks2014}, see Fig.~\ref{fig:cell}(a). Mechanically, the cell is composed of two different parts A and B of different spring constants which are then connected by the sample in mechanical series, as shown in Fig.~\ref{fig:cell}(b). Being in series, all parts along with the sample experience the same force but different stresses. This force on the cell is  measured by a force sensor placed at the end of device, which consists of four strain gauges mounted in a Wheatstone bridge configuration. The cell is controlled by a feedback loop written in Python. 

\subsection{Sample preparation}
The single crystal, single domain of 5$\%$ MgO-doped LiNbO$_3$ used in this work were obatined from \textquotedblleft Yamaju Ceramics Co., Ltd.\textquotedblright. The domains in the crystal were subsequently written by the UV-assited poling method \cite{Muller2003}. The samples were later cut to narrow bars, see Fig. \ref{fig:domain_concept}(e), such that the sample LNO-01 and LNO-02 have the cross sections of 310 $\upmu$m $\times$ 200 $\upmu$m (z $\times$ y) and 370 $\upmu$m $\times$ 200 $\upmu$m (z $\times$ x), respectively. Afterwards fabricated domain structures were treated to increase the conductivity by the procedure developed by Godau et al. \cite{Godau2017}. In the enhancement procedure chromium electrodes were used to apply the -500V DC voltage to 200 $\upmu$m thick LiNbO$_3$ crystal. With this the DWs moved and the inclination angle of DWs increased by $\sim$ 1$^\circ$ \cite{Kampfe2014}, hence the current increased by a factor of 10$^4$ (at -10V) compared to as-poled DWs (see Fig. S2, S3 of  supplement and section on angle calculation).

\subsection{Scanning force microscopy(PFM and cAFM)}

The measurements were performed on a NX10 scanning probe microscope from \textquotedblleft Park Systems Corp.\textquotedblright. PFM \cite{Soergel2011} on the samples was performed using pure platinum tips model RMN-25PT300B (free resonance frequency, $f_{free}$ $\approx$ 20 kHz) as a top contact, while applying the external mechanical stress. For PFM, an alternating voltage of 20 V$_{p-p}$ at a frequency smaller than contact resonance frequency ( $f_{cont}$ $\approx$ 41 kHz) was applied to the cantilever. 

Conductive atomic force microscopy (cAFM) was also performed using the same RMN tips. The sample in this case was kept at -10V while the cantilever and the stress cell were kept at ground, see Fig.~\ref{fig:cell}(c).

\section{Conclusion}\label{sec13}
In summary, we demonstrate that uniaxial compressive or tensile stress along the x and y- axes can be used to purposely tune the conductivity of ferroelectric DWs in LiNbO$_3$. Here, applying uniaxial stress to a sample results in a change of screening charges on the domain boundary due to the piezoelectric effect. Depending on the relative orientations of the stress and the DW respectively, a DW can deliberately be configured h2h or t2t, for example. Based on this, our results show that the conductivity is approximately proportional to the amount of induced negative screening charges, i.e. i(h2h) configurations, while induced positively charged walls, i.e. i(t2t) configurations, show a rapid decrease in overall conductivity down to bulk conductivity. This  observation hints towards electron polarons, rather than hole polarons, that are the main contributor to the conductivity of DWs in LiNbO$_3$ \cite{Eliseev2011}. The same behavior was observed when measurements were repeated many times over a period of two months (see supplement section S3.2). Our measurements were performed on a stress cell that allows us to control uniaxial stress independent of a substrate or temperature and, therefore, offering large flexibility for studying different geometries or materials. In conclusion, with uniaxial stress we can gain a directional control of DW conductivity potentially allowing novel applications  such as stress or strain based nanosensors, as well as providing fundamental insights into the properties of DWs.

\section{Acknowledgments}
We express our great gratitude for financial support by the Deutsche Forschungsgemeinschaft (DFG) through joint DFG-ANR project \textquotedblleft TOPELEC \textquotedblright (Nos. EN 434/41-1 and ANR-18-J.CE92-0052-1), the CRC1415 (ID: 417590517), the FOR5044 (ID: 426703838) www.FOR5044.de, as well as by the W{\"u}rzburg-Dresden Cluster of Excellence on \textquotedblleft  Complexity and Topology in Quantum Matter\textquotedblright - ct.qmat (EXC 2147; ID: 39085490). We also acknowledge the excellent support by the Light Microscopy Facility, a Core Facility of the CMCB Technology Platform at TU Dresden, where the SHG analysis was performed and Park Systems, on which the cAFM measurements were performed. Furthermore, we thank D. Bieberstein and T. Gemming from IFW Dresden for assistance with dicing of the wafers, Dr. Elke Beyreuther for valuable discussions and Ahmed Samir Lotfy for his lab assistance to this work.
\nocite{*}
\bibliography{ms}

\begin{thebibliography}{37}%
\makeatletter
\providecommand \@ifxundefined [1]{%
 \@ifx{#1\undefined}
}%
\providecommand \@ifnum [1]{%
 \ifnum #1\expandafter \@firstoftwo
 \else \expandafter \@secondoftwo
 \fi
}%
\providecommand \@ifx [1]{%
 \ifx #1\expandafter \@firstoftwo
 \else \expandafter \@secondoftwo
 \fi
}%
\providecommand \natexlab [1]{#1}%
\providecommand \enquote  [1]{``#1''}%
\providecommand \bibnamefont  [1]{#1}%
\providecommand \bibfnamefont [1]{#1}%
\providecommand \citenamefont [1]{#1}%
\providecommand \href@noop [0]{\@secondoftwo}%
\providecommand \href [0]{\begingroup \@sanitize@url \@href}%
\providecommand \@href[1]{\@@startlink{#1}\@@href}%
\providecommand \@@href[1]{\endgroup#1\@@endlink}%
\providecommand \@sanitize@url [0]{\catcode `\\12\catcode `\$12\catcode
  `\&12\catcode `\#12\catcode `\^12\catcode `\_12\catcode `\%12\relax}%
\providecommand \@@startlink[1]{}%
\providecommand \@@endlink[0]{}%
\providecommand \url  [0]{\begingroup\@sanitize@url \@url }%
\providecommand \@url [1]{\endgroup\@href {#1}{\urlprefix }}%
\providecommand \urlprefix  [0]{URL }%
\providecommand \Eprint [0]{\href }%
\providecommand \doibase [0]{https://doi.org/}%
\providecommand \selectlanguage [0]{\@gobble}%
\providecommand \bibinfo  [0]{\@secondoftwo}%
\providecommand \bibfield  [0]{\@secondoftwo}%
\providecommand \translation [1]{[#1]}%
\providecommand \BibitemOpen [0]{}%
\providecommand \bibitemStop [0]{}%
\providecommand \bibitemNoStop [0]{.\EOS\space}%
\providecommand \EOS [0]{\spacefactor3000\relax}%
\providecommand \BibitemShut  [1]{\csname bibitem#1\endcsname}%
\let\auto@bib@innerbib\@empty
\bibitem [{\citenamefont {Kirbus}\ \emph
  {et~al.}(2019{\natexlab{a}})\citenamefont {Kirbus}, \citenamefont {Godau},
  \citenamefont {Wehmeier}, \citenamefont {Beccard}, \citenamefont
  {Beyreuther}, \citenamefont {Hau{\ss}mann},\ and\ \citenamefont
  {Eng}}]{Kirbus2019}%
  \BibitemOpen
  \bibfield  {author} {\bibinfo {author} {\bibfnamefont {B.}~\bibnamefont
  {Kirbus}}, \bibinfo {author} {\bibfnamefont {C.}~\bibnamefont {Godau}},
  \bibinfo {author} {\bibfnamefont {L.}~\bibnamefont {Wehmeier}}, \bibinfo
  {author} {\bibfnamefont {H.}~\bibnamefont {Beccard}}, \bibinfo {author}
  {\bibfnamefont {E.}~\bibnamefont {Beyreuther}}, \bibinfo {author}
  {\bibfnamefont {A.}~\bibnamefont {Hau{\ss}mann}},\ and\ \bibinfo {author}
  {\bibfnamefont {L.~M.}\ \bibnamefont {Eng}},\ }\bibfield  {title} {\bibinfo
  {title} {{Real-Time 3D Imaging of Nanoscale Ferroelectric Domain Wall
  Dynamics in Lithium Niobate Single Crystals under Electric Stimuli:
  Implications for Domain-Wall-Based Nanoelectronic Devices}},\ }\href
  {https://doi.org/10.1021/acsanm.9b01240} {\bibfield  {journal} {\bibinfo
  {journal} {ACS Applied Nano Materials}\ }\textbf {\bibinfo {volume} {2}},\
  \bibinfo {pages} {5787} (\bibinfo {year} {2019}{\natexlab{a}})}\BibitemShut
  {NoStop}%
\bibitem [{\citenamefont {Sharma}\ \emph {et~al.}(2017)\citenamefont {Sharma},
  \citenamefont {Zhang}, \citenamefont {Sando}, \citenamefont {Lei},
  \citenamefont {Liu}, \citenamefont {Li}, \citenamefont {Nagarajan},\ and\
  \citenamefont {Seidel}}]{Sharma2017a}%
  \BibitemOpen
  \bibfield  {author} {\bibinfo {author} {\bibfnamefont {P.}~\bibnamefont
  {Sharma}}, \bibinfo {author} {\bibfnamefont {Q.}~\bibnamefont {Zhang}},
  \bibinfo {author} {\bibfnamefont {D.}~\bibnamefont {Sando}}, \bibinfo
  {author} {\bibfnamefont {C.~H.}\ \bibnamefont {Lei}}, \bibinfo {author}
  {\bibfnamefont {Y.}~\bibnamefont {Liu}}, \bibinfo {author} {\bibfnamefont
  {J.}~\bibnamefont {Li}}, \bibinfo {author} {\bibfnamefont {V.}~\bibnamefont
  {Nagarajan}},\ and\ \bibinfo {author} {\bibfnamefont {J.}~\bibnamefont
  {Seidel}},\ }\bibfield  {title} {\bibinfo {title} {{Nonvolatile ferroelectric
  domain wall memory}},\ }\href {https://doi.org/10.1126/sciadv.1700512}
  {\bibfield  {journal} {\bibinfo  {journal} {Science Advances}\ }\textbf
  {\bibinfo {volume} {3}},\ \bibinfo {pages} {1} (\bibinfo {year}
  {2017})}\BibitemShut {NoStop}%
\bibitem [{\citenamefont {Zhao}\ \emph {et~al.}(2020)\citenamefont {Zhao},
  \citenamefont {Ma}, \citenamefont {R{\"{u}}sing},\ and\ \citenamefont
  {Mookherjea}}]{Zhao2020}%
  \BibitemOpen
  \bibfield  {author} {\bibinfo {author} {\bibfnamefont {J.}~\bibnamefont
  {Zhao}}, \bibinfo {author} {\bibfnamefont {C.}~\bibnamefont {Ma}}, \bibinfo
  {author} {\bibfnamefont {M.}~\bibnamefont {R{\"{u}}sing}},\ and\ \bibinfo
  {author} {\bibfnamefont {S.}~\bibnamefont {Mookherjea}},\ }\bibfield  {title}
  {\bibinfo {title} {{High Quality Entangled Photon Pair Generation in
  Periodically Poled Thin-Film Lithium Niobate Waveguides}},\ }\href
  {https://doi.org/10.1103/PhysRevLett.124.163603} {\bibfield  {journal}
  {\bibinfo  {journal} {Physical Review Letters}\ }\textbf {\bibinfo {volume}
  {124}},\ \bibinfo {pages} {163603} (\bibinfo {year} {2020})}\BibitemShut
  {NoStop}%
\bibitem [{\citenamefont {K{\"{a}}mpfe}\ \emph {et~al.}(2020)\citenamefont
  {K{\"{a}}mpfe}, \citenamefont {Wang}, \citenamefont {Hau{\ss}mann},
  \citenamefont {Chen},\ and\ \citenamefont {Eng}}]{Kampfe2020b}%
  \BibitemOpen
  \bibfield  {author} {\bibinfo {author} {\bibfnamefont {T.}~\bibnamefont
  {K{\"{a}}mpfe}}, \bibinfo {author} {\bibfnamefont {B.}~\bibnamefont {Wang}},
  \bibinfo {author} {\bibfnamefont {A.}~\bibnamefont {Hau{\ss}mann}}, \bibinfo
  {author} {\bibfnamefont {L.-Q.}\ \bibnamefont {Chen}},\ and\ \bibinfo
  {author} {\bibfnamefont {L.~M.}\ \bibnamefont {Eng}},\ }\bibfield  {title}
  {\bibinfo {title} {{Tunable Non-Volatile Memory by Conductive Ferroelectric
  Domain Walls in Lithium Niobate Thin Films}},\ }\href
  {https://doi.org/10.3390/cryst10090804} {\bibfield  {journal} {\bibinfo
  {journal} {Crystals}\ }\textbf {\bibinfo {volume} {10}},\ \bibinfo {pages}
  {804} (\bibinfo {year} {2020})}\BibitemShut {NoStop}%
\bibitem [{\citenamefont {Jiang}\ \emph {et~al.}(2018)\citenamefont {Jiang},
  \citenamefont {Bai}, \citenamefont {Chen}, \citenamefont {He}, \citenamefont
  {Zhang}, \citenamefont {Zhang}, \citenamefont {Shi}, \citenamefont {Park},
  \citenamefont {Scott}, \citenamefont {Hwang},\ and\ \citenamefont
  {Jiang}}]{Jiang2018}%
  \BibitemOpen
  \bibfield  {author} {\bibinfo {author} {\bibfnamefont {J.}~\bibnamefont
  {Jiang}}, \bibinfo {author} {\bibfnamefont {Z.~L.}\ \bibnamefont {Bai}},
  \bibinfo {author} {\bibfnamefont {Z.~H.}\ \bibnamefont {Chen}}, \bibinfo
  {author} {\bibfnamefont {L.}~\bibnamefont {He}}, \bibinfo {author}
  {\bibfnamefont {D.~W.}\ \bibnamefont {Zhang}}, \bibinfo {author}
  {\bibfnamefont {Q.~H.}\ \bibnamefont {Zhang}}, \bibinfo {author}
  {\bibfnamefont {J.~A.}\ \bibnamefont {Shi}}, \bibinfo {author} {\bibfnamefont
  {M.~H.}\ \bibnamefont {Park}}, \bibinfo {author} {\bibfnamefont {J.~F.}\
  \bibnamefont {Scott}}, \bibinfo {author} {\bibfnamefont {C.~S.}\ \bibnamefont
  {Hwang}},\ and\ \bibinfo {author} {\bibfnamefont {A.~Q.}\ \bibnamefont
  {Jiang}},\ }\bibfield  {title} {\bibinfo {title} {{Temporary formation of
  highly conducting domain walls for non-destructive read-out of ferroelectric
  domain-wall resistance switching memories}},\ }\href
  {https://doi.org/10.1038/NMAT5028} {\bibfield  {journal} {\bibinfo  {journal}
  {Nature Materials}\ }\textbf {\bibinfo {volume} {17}},\ \bibinfo {pages} {49}
  (\bibinfo {year} {2018})}\BibitemShut {NoStop}%
\bibitem [{\citenamefont {Li}\ \emph {et~al.}(2016)\citenamefont {Li},
  \citenamefont {Britson}, \citenamefont {Jokisaari}, \citenamefont {Zhang},
  \citenamefont {Adamo}, \citenamefont {Melville}, \citenamefont {Schlom},
  \citenamefont {Chen},\ and\ \citenamefont {Pan}}]{Li2016}%
  \BibitemOpen
  \bibfield  {author} {\bibinfo {author} {\bibfnamefont {L.}~\bibnamefont
  {Li}}, \bibinfo {author} {\bibfnamefont {J.}~\bibnamefont {Britson}},
  \bibinfo {author} {\bibfnamefont {J.~R.}\ \bibnamefont {Jokisaari}}, \bibinfo
  {author} {\bibfnamefont {Y.}~\bibnamefont {Zhang}}, \bibinfo {author}
  {\bibfnamefont {C.}~\bibnamefont {Adamo}}, \bibinfo {author} {\bibfnamefont
  {A.}~\bibnamefont {Melville}}, \bibinfo {author} {\bibfnamefont {D.~G.}\
  \bibnamefont {Schlom}}, \bibinfo {author} {\bibfnamefont {L.~Q.}\
  \bibnamefont {Chen}},\ and\ \bibinfo {author} {\bibfnamefont
  {X.}~\bibnamefont {Pan}},\ }\bibfield  {title} {\bibinfo {title} {{Giant
  Resistive Switching via Control of Ferroelectric Charged Domain Walls}},\
  }\href {https://doi.org/10.1002/adma.201600160} {\bibfield  {journal}
  {\bibinfo  {journal} {Advanced Materials}\ }\textbf {\bibinfo {volume}
  {28}},\ \bibinfo {pages} {6574} (\bibinfo {year} {2016})}\BibitemShut
  {NoStop}%
\bibitem [{\citenamefont {Gruverman}\ \emph {et~al.}(2008)\citenamefont
  {Gruverman}, \citenamefont {Wu},\ and\ \citenamefont
  {Scott}}]{Gruverman2008}%
  \BibitemOpen
  \bibfield  {author} {\bibinfo {author} {\bibfnamefont {A.}~\bibnamefont
  {Gruverman}}, \bibinfo {author} {\bibfnamefont {D.}~\bibnamefont {Wu}},\ and\
  \bibinfo {author} {\bibfnamefont {J.~F.}\ \bibnamefont {Scott}},\ }\bibfield
  {title} {\bibinfo {title} {{Piezoresponse force microscopy studies of
  switching behavior of ferroelectric capacitors on a 100-ns time scale}},\
  }\href {https://doi.org/10.1103/PhysRevLett.100.097601} {\bibfield  {journal}
  {\bibinfo  {journal} {Physical Review Letters}\ }\textbf {\bibinfo {volume}
  {100}},\ \bibinfo {pages} {3} (\bibinfo {year} {2008})}\BibitemShut {NoStop}%
\bibitem [{\citenamefont {Catalan}\ \emph {et~al.}(2012)\citenamefont
  {Catalan}, \citenamefont {Seidel}, \citenamefont {Ramesh},\ and\
  \citenamefont {Scott}}]{Catalan2012}%
  \BibitemOpen
  \bibfield  {author} {\bibinfo {author} {\bibfnamefont {G.}~\bibnamefont
  {Catalan}}, \bibinfo {author} {\bibfnamefont {J.}~\bibnamefont {Seidel}},
  \bibinfo {author} {\bibfnamefont {R.}~\bibnamefont {Ramesh}},\ and\ \bibinfo
  {author} {\bibfnamefont {J.~F.}\ \bibnamefont {Scott}},\ }\bibfield  {title}
  {\bibinfo {title} {{Domain wall nanoelectronics}},\ }\href
  {https://doi.org/10.1103/RevModPhys.84.119} {\bibfield  {journal} {\bibinfo
  {journal} {Reviews of Modern Physics}\ }\textbf {\bibinfo {volume} {84}},\
  \bibinfo {pages} {119} (\bibinfo {year} {2012})}\BibitemShut {NoStop}%
\bibitem [{\citenamefont {{Seidel Editor}}\ and\ \citenamefont
  {Walls}(2016)}]{SeidelEditora}%
  \BibitemOpen
  \bibfield  {author} {\bibinfo {author} {\bibfnamefont {J.}~\bibnamefont
  {{Seidel Editor}}}\ and\ \bibinfo {author} {\bibfnamefont {D.}~\bibnamefont
  {Walls}},\ }\href {https://doi.org/10.1007/978-3-319-25301-5} {\emph
  {\bibinfo {title} {{Topological Structures in Ferroic Materials}}}},\ edited
  by\ \bibinfo {editor} {\bibfnamefont {J.}~\bibnamefont {Seidel}},\ \bibinfo
  {series} {Springer Series in Materials Science}, Vol.\ \bibinfo {volume}
  {228}\ (\bibinfo  {publisher} {Springer International Publishing},\ \bibinfo
  {address} {Cham},\ \bibinfo {year} {2016})\ pp.\ \bibinfo {pages}
  {103--138}\BibitemShut {NoStop}%
\bibitem [{\citenamefont {Sluka}\ \emph {et~al.}(2012)\citenamefont {Sluka},
  \citenamefont {Tagantsev}, \citenamefont {Damjanovic}, \citenamefont
  {Gureev},\ and\ \citenamefont {Setter}}]{Sluka2012}%
  \BibitemOpen
  \bibfield  {author} {\bibinfo {author} {\bibfnamefont {T.}~\bibnamefont
  {Sluka}}, \bibinfo {author} {\bibfnamefont {A.~K.}\ \bibnamefont
  {Tagantsev}}, \bibinfo {author} {\bibfnamefont {D.}~\bibnamefont
  {Damjanovic}}, \bibinfo {author} {\bibfnamefont {M.}~\bibnamefont {Gureev}},\
  and\ \bibinfo {author} {\bibfnamefont {N.}~\bibnamefont {Setter}},\
  }\bibfield  {title} {\bibinfo {title} {{Enhanced electromechanical response
  of ferroelectrics due to charged domain walls}},\ }\href
  {https://doi.org/10.1038/ncomms1751} {\bibfield  {journal} {\bibinfo
  {journal} {Nature Communications}\ }\textbf {\bibinfo {volume} {3}},\
  \bibinfo {pages} {748} (\bibinfo {year} {2012})}\BibitemShut {NoStop}%
\bibitem [{\citenamefont {Condurache}\ \emph {et~al.}(2021)\citenamefont
  {Condurache}, \citenamefont {Dra{\v{z}}i{\'{c}}}, \citenamefont {Sakamoto},
  \citenamefont {Rojac},\ and\ \citenamefont {Ben{\v{c}}an}}]{Condurache2021}%
  \BibitemOpen
  \bibfield  {author} {\bibinfo {author} {\bibfnamefont {O.}~\bibnamefont
  {Condurache}}, \bibinfo {author} {\bibfnamefont {G.}~\bibnamefont
  {Dra{\v{z}}i{\'{c}}}}, \bibinfo {author} {\bibfnamefont {N.}~\bibnamefont
  {Sakamoto}}, \bibinfo {author} {\bibfnamefont {T.}~\bibnamefont {Rojac}},\
  and\ \bibinfo {author} {\bibfnamefont {A.}~\bibnamefont {Ben{\v{c}}an}},\
  }\bibfield  {title} {\bibinfo {title} {{Atomically resolved structure of
  step-like uncharged and charged domain walls in polycrystalline BiFeO 3}},\
  }\href {https://doi.org/10.1063/5.0034699} {\bibfield  {journal} {\bibinfo
  {journal} {Journal of Applied Physics}\ }\textbf {\bibinfo {volume} {129}},\
  \bibinfo {pages} {054102} (\bibinfo {year} {2021})}\BibitemShut {NoStop}%
\bibitem [{\citenamefont {Rojac}\ \emph {et~al.}(2017)\citenamefont {Rojac},
  \citenamefont {Bencan}, \citenamefont {Drazic}, \citenamefont {Sakamoto},
  \citenamefont {Ursic}, \citenamefont {Jancar}, \citenamefont {Tavcar},
  \citenamefont {Makarovic}, \citenamefont {Walker}, \citenamefont {Malic},\
  and\ \citenamefont {Damjanovic}}]{Rojac2017}%
  \BibitemOpen
  \bibfield  {author} {\bibinfo {author} {\bibfnamefont {T.}~\bibnamefont
  {Rojac}}, \bibinfo {author} {\bibfnamefont {A.}~\bibnamefont {Bencan}},
  \bibinfo {author} {\bibfnamefont {G.}~\bibnamefont {Drazic}}, \bibinfo
  {author} {\bibfnamefont {N.}~\bibnamefont {Sakamoto}}, \bibinfo {author}
  {\bibfnamefont {H.}~\bibnamefont {Ursic}}, \bibinfo {author} {\bibfnamefont
  {B.}~\bibnamefont {Jancar}}, \bibinfo {author} {\bibfnamefont
  {G.}~\bibnamefont {Tavcar}}, \bibinfo {author} {\bibfnamefont
  {M.}~\bibnamefont {Makarovic}}, \bibinfo {author} {\bibfnamefont
  {J.}~\bibnamefont {Walker}}, \bibinfo {author} {\bibfnamefont
  {B.}~\bibnamefont {Malic}},\ and\ \bibinfo {author} {\bibfnamefont
  {D.}~\bibnamefont {Damjanovic}},\ }\bibfield  {title} {\bibinfo {title}
  {{Domain-wall conduction in ferroelectric BiFeO 3 controlled by accumulation
  of charged defects}},\ }\href {https://doi.org/10.1038/nmat4799} {\bibfield
  {journal} {\bibinfo  {journal} {Nature Materials}\ }\textbf {\bibinfo
  {volume} {16}},\ \bibinfo {pages} {322} (\bibinfo {year} {2017})}\BibitemShut
  {NoStop}%
\bibitem [{\citenamefont {Stolichnov}\ \emph {et~al.}(2015)\citenamefont
  {Stolichnov}, \citenamefont {Feigl}, \citenamefont {McGilly}, \citenamefont
  {Sluka}, \citenamefont {Wei}, \citenamefont {Colla}, \citenamefont
  {Crassous}, \citenamefont {Shapovalov}, \citenamefont {Yudin}, \citenamefont
  {Tagantsev},\ and\ \citenamefont {Setter}}]{Stolichnov2015}%
  \BibitemOpen
  \bibfield  {author} {\bibinfo {author} {\bibfnamefont {I.}~\bibnamefont
  {Stolichnov}}, \bibinfo {author} {\bibfnamefont {L.}~\bibnamefont {Feigl}},
  \bibinfo {author} {\bibfnamefont {L.~J.}\ \bibnamefont {McGilly}}, \bibinfo
  {author} {\bibfnamefont {T.}~\bibnamefont {Sluka}}, \bibinfo {author}
  {\bibfnamefont {X.-K.}\ \bibnamefont {Wei}}, \bibinfo {author} {\bibfnamefont
  {E.}~\bibnamefont {Colla}}, \bibinfo {author} {\bibfnamefont
  {A.}~\bibnamefont {Crassous}}, \bibinfo {author} {\bibfnamefont
  {K.}~\bibnamefont {Shapovalov}}, \bibinfo {author} {\bibfnamefont
  {P.}~\bibnamefont {Yudin}}, \bibinfo {author} {\bibfnamefont {A.~K.}\
  \bibnamefont {Tagantsev}},\ and\ \bibinfo {author} {\bibfnamefont
  {N.}~\bibnamefont {Setter}},\ }\bibfield  {title} {\bibinfo {title} {{Bent
  Ferroelectric Domain Walls as Reconfigurable Metallic-Like Channels}},\
  }\href {https://doi.org/10.1021/acs.nanolett.5b03450} {\bibfield  {journal}
  {\bibinfo  {journal} {Nano Letters}\ }\textbf {\bibinfo {volume} {15}},\
  \bibinfo {pages} {8049} (\bibinfo {year} {2015})}\BibitemShut {NoStop}%
\bibitem [{\citenamefont {Sluka}\ \emph {et~al.}(2013)\citenamefont {Sluka},
  \citenamefont {Tagantsev}, \citenamefont {Bednyakov},\ and\ \citenamefont
  {Setter}}]{Sluka2013}%
  \BibitemOpen
  \bibfield  {author} {\bibinfo {author} {\bibfnamefont {T.}~\bibnamefont
  {Sluka}}, \bibinfo {author} {\bibfnamefont {A.~K.}\ \bibnamefont
  {Tagantsev}}, \bibinfo {author} {\bibfnamefont {P.}~\bibnamefont
  {Bednyakov}},\ and\ \bibinfo {author} {\bibfnamefont {N.}~\bibnamefont
  {Setter}},\ }\bibfield  {title} {\bibinfo {title} {{Free-electron gas at
  charged domain walls in insulating BaTiO3}},\ }\href
  {https://doi.org/10.1038/ncomms2839} {\bibfield  {journal} {\bibinfo
  {journal} {Nature Communications}\ }\textbf {\bibinfo {volume} {4}},\
  \bibinfo {pages} {1806} (\bibinfo {year} {2013})}\BibitemShut {NoStop}%
\bibitem [{\citenamefont {Wu}\ \emph {et~al.}(2012)\citenamefont {Wu},
  \citenamefont {Horibe}, \citenamefont {Lee}, \citenamefont {Cheong},\ and\
  \citenamefont {Guest}}]{Wu2012}%
  \BibitemOpen
  \bibfield  {author} {\bibinfo {author} {\bibfnamefont {W.}~\bibnamefont
  {Wu}}, \bibinfo {author} {\bibfnamefont {Y.}~\bibnamefont {Horibe}}, \bibinfo
  {author} {\bibfnamefont {N.}~\bibnamefont {Lee}}, \bibinfo {author}
  {\bibfnamefont {S.-W.}\ \bibnamefont {Cheong}},\ and\ \bibinfo {author}
  {\bibfnamefont {J.~R.}\ \bibnamefont {Guest}},\ }\bibfield  {title} {\bibinfo
  {title} {{Conduction of Topologically Protected Charged Ferroelectric Domain
  Walls}},\ }\href {https://doi.org/10.1103/PhysRevLett.108.077203} {\bibfield
  {journal} {\bibinfo  {journal} {Physical Review Letters}\ }\textbf {\bibinfo
  {volume} {108}},\ \bibinfo {pages} {077203} (\bibinfo {year}
  {2012})}\BibitemShut {NoStop}%
\bibitem [{\citenamefont {Gonnissen}\ \emph {et~al.}(2016)\citenamefont
  {Gonnissen}, \citenamefont {Batuk}, \citenamefont {Nataf}, \citenamefont
  {Jones}, \citenamefont {Abakumov}, \citenamefont {{Van Aert}}, \citenamefont
  {Schryvers},\ and\ \citenamefont {Salje}}]{Gonnissen2016}%
  \BibitemOpen
  \bibfield  {author} {\bibinfo {author} {\bibfnamefont {J.}~\bibnamefont
  {Gonnissen}}, \bibinfo {author} {\bibfnamefont {D.}~\bibnamefont {Batuk}},
  \bibinfo {author} {\bibfnamefont {G.~F.}\ \bibnamefont {Nataf}}, \bibinfo
  {author} {\bibfnamefont {L.}~\bibnamefont {Jones}}, \bibinfo {author}
  {\bibfnamefont {A.~M.}\ \bibnamefont {Abakumov}}, \bibinfo {author}
  {\bibfnamefont {S.}~\bibnamefont {{Van Aert}}}, \bibinfo {author}
  {\bibfnamefont {D.}~\bibnamefont {Schryvers}},\ and\ \bibinfo {author}
  {\bibfnamefont {E.~K.~H.}\ \bibnamefont {Salje}},\ }\bibfield  {title}
  {\bibinfo {title} {{Direct Observation of Ferroelectric Domain Walls in LiNbO
  3 : Wall-Meanders, Kinks, and Local Electric Charges}},\ }\href
  {https://doi.org/10.1002/adfm.201603489} {\bibfield  {journal} {\bibinfo
  {journal} {Advanced Functional Materials}\ }\textbf {\bibinfo {volume}
  {26}},\ \bibinfo {pages} {7599} (\bibinfo {year} {2016})}\BibitemShut
  {NoStop}%
\bibitem [{\citenamefont {Werner}\ \emph {et~al.}(2017)\citenamefont {Werner},
  \citenamefont {Herr}, \citenamefont {Buse}, \citenamefont {Sturman},
  \citenamefont {Soergel}, \citenamefont {Razzaghi},\ and\ \citenamefont
  {Breunig}}]{Werner2017}%
  \BibitemOpen
  \bibfield  {author} {\bibinfo {author} {\bibfnamefont {C.~S.}\ \bibnamefont
  {Werner}}, \bibinfo {author} {\bibfnamefont {S.~J.}\ \bibnamefont {Herr}},
  \bibinfo {author} {\bibfnamefont {K.}~\bibnamefont {Buse}}, \bibinfo {author}
  {\bibfnamefont {B.}~\bibnamefont {Sturman}}, \bibinfo {author} {\bibfnamefont
  {E.}~\bibnamefont {Soergel}}, \bibinfo {author} {\bibfnamefont
  {C.}~\bibnamefont {Razzaghi}},\ and\ \bibinfo {author} {\bibfnamefont
  {I.}~\bibnamefont {Breunig}},\ }\bibfield  {title} {\bibinfo {title} {{Large
  and accessible conductivity of charged domain walls in lithium niobate}},\
  }\href {https://doi.org/10.1038/s41598-017-09703-2} {\bibfield  {journal}
  {\bibinfo  {journal} {Scientific Reports}\ }\textbf {\bibinfo {volume} {7}},\
  \bibinfo {pages} {9862} (\bibinfo {year} {2017})}\BibitemShut {NoStop}%
\bibitem [{\citenamefont {Schr{\"{o}}der}\ \emph {et~al.}(2012)\citenamefont
  {Schr{\"{o}}der}, \citenamefont {Hau{\ss}mann}, \citenamefont {Thiessen},
  \citenamefont {Soergel}, \citenamefont {Woike},\ and\ \citenamefont
  {Eng}}]{Schroder2012}%
  \BibitemOpen
  \bibfield  {author} {\bibinfo {author} {\bibfnamefont {M.}~\bibnamefont
  {Schr{\"{o}}der}}, \bibinfo {author} {\bibfnamefont {A.}~\bibnamefont
  {Hau{\ss}mann}}, \bibinfo {author} {\bibfnamefont {A.}~\bibnamefont
  {Thiessen}}, \bibinfo {author} {\bibfnamefont {E.}~\bibnamefont {Soergel}},
  \bibinfo {author} {\bibfnamefont {T.}~\bibnamefont {Woike}},\ and\ \bibinfo
  {author} {\bibfnamefont {L.~M.}\ \bibnamefont {Eng}},\ }\bibfield  {title}
  {\bibinfo {title} {{Conducting Domain Walls in Lithium Niobate Single
  Crystals}},\ }\href {https://doi.org/10.1002/adfm.201201174} {\bibfield
  {journal} {\bibinfo  {journal} {Advanced Functional Materials}\ }\textbf
  {\bibinfo {volume} {22}},\ \bibinfo {pages} {3936} (\bibinfo {year}
  {2012})}\BibitemShut {NoStop}%
\bibitem [{\citenamefont {Godau}\ \emph {et~al.}(2017)\citenamefont {Godau},
  \citenamefont {K{\"{a}}mpfe}, \citenamefont {Thiessen}, \citenamefont {Eng},\
  and\ \citenamefont {Hau{\ss}mann}}]{Godau2017}%
  \BibitemOpen
  \bibfield  {author} {\bibinfo {author} {\bibfnamefont {C.}~\bibnamefont
  {Godau}}, \bibinfo {author} {\bibfnamefont {T.}~\bibnamefont {K{\"{a}}mpfe}},
  \bibinfo {author} {\bibfnamefont {A.}~\bibnamefont {Thiessen}}, \bibinfo
  {author} {\bibfnamefont {L.~M.}\ \bibnamefont {Eng}},\ and\ \bibinfo {author}
  {\bibfnamefont {A.}~\bibnamefont {Hau{\ss}mann}},\ }\bibfield  {title}
  {\bibinfo {title} {{Enhancing the Domain Wall Conductivity in Lithium Niobate
  Single Crystals}},\ }\href {https://doi.org/10.1021/acsnano.7b01199}
  {\bibfield  {journal} {\bibinfo  {journal} {ACS Nano}\ }\textbf {\bibinfo
  {volume} {11}},\ \bibinfo {pages} {4816} (\bibinfo {year}
  {2017})}\BibitemShut {NoStop}%
\bibitem [{\citenamefont {Lu}\ \emph {et~al.}(2019)\citenamefont {Lu},
  \citenamefont {Tan}, \citenamefont {McConville}, \citenamefont {Ahmadi},
  \citenamefont {Wang}, \citenamefont {Conroy}, \citenamefont {Moore},
  \citenamefont {Bangert}, \citenamefont {Shield}, \citenamefont {Chen},
  \citenamefont {Gregg},\ and\ \citenamefont {Gruverman}}]{Lu2019}%
  \BibitemOpen
  \bibfield  {author} {\bibinfo {author} {\bibfnamefont {H.}~\bibnamefont
  {Lu}}, \bibinfo {author} {\bibfnamefont {Y.}~\bibnamefont {Tan}}, \bibinfo
  {author} {\bibfnamefont {J.~P.~V.}\ \bibnamefont {McConville}}, \bibinfo
  {author} {\bibfnamefont {Z.}~\bibnamefont {Ahmadi}}, \bibinfo {author}
  {\bibfnamefont {B.}~\bibnamefont {Wang}}, \bibinfo {author} {\bibfnamefont
  {M.}~\bibnamefont {Conroy}}, \bibinfo {author} {\bibfnamefont
  {K.}~\bibnamefont {Moore}}, \bibinfo {author} {\bibfnamefont
  {U.}~\bibnamefont {Bangert}}, \bibinfo {author} {\bibfnamefont {J.~E.}\
  \bibnamefont {Shield}}, \bibinfo {author} {\bibfnamefont {L.}~\bibnamefont
  {Chen}}, \bibinfo {author} {\bibfnamefont {J.~M.}\ \bibnamefont {Gregg}},\
  and\ \bibinfo {author} {\bibfnamefont {A.}~\bibnamefont {Gruverman}},\
  }\bibfield  {title} {\bibinfo {title} {{Electrical Tunability of Domain Wall
  Conductivity in LiNbO 3 Thin Films}},\ }\href
  {https://doi.org/10.1002/adma.201902890} {\bibfield  {journal} {\bibinfo
  {journal} {Advanced Materials}\ }\textbf {\bibinfo {volume} {31}},\ \bibinfo
  {pages} {1902890} (\bibinfo {year} {2019})}\BibitemShut {NoStop}%
\bibitem [{\citenamefont {Chen}\ \emph {et~al.}(2012)\citenamefont {Chen},
  \citenamefont {Zou}, \citenamefont {Ren}, \citenamefont {You}, \citenamefont
  {Huang}, \citenamefont {Yang}, \citenamefont {Yang}, \citenamefont {Wang},
  \citenamefont {Sritharan}, \citenamefont {Bellaiche},\ and\ \citenamefont
  {Chen}}]{Chen2012}%
  \BibitemOpen
  \bibfield  {author} {\bibinfo {author} {\bibfnamefont {Z.}~\bibnamefont
  {Chen}}, \bibinfo {author} {\bibfnamefont {X.}~\bibnamefont {Zou}}, \bibinfo
  {author} {\bibfnamefont {W.}~\bibnamefont {Ren}}, \bibinfo {author}
  {\bibfnamefont {L.}~\bibnamefont {You}}, \bibinfo {author} {\bibfnamefont
  {C.}~\bibnamefont {Huang}}, \bibinfo {author} {\bibfnamefont
  {Y.}~\bibnamefont {Yang}}, \bibinfo {author} {\bibfnamefont {P.}~\bibnamefont
  {Yang}}, \bibinfo {author} {\bibfnamefont {J.}~\bibnamefont {Wang}}, \bibinfo
  {author} {\bibfnamefont {T.}~\bibnamefont {Sritharan}}, \bibinfo {author}
  {\bibfnamefont {L.}~\bibnamefont {Bellaiche}},\ and\ \bibinfo {author}
  {\bibfnamefont {L.}~\bibnamefont {Chen}},\ }\bibfield  {title} {\bibinfo
  {title} {{Study of strain effect on in-plane polarization in epitaxial BiFeO
  3 thin films using planar electrodes}},\ }\href
  {https://doi.org/10.1103/PhysRevB.86.235125} {\bibfield  {journal} {\bibinfo
  {journal} {Physical Review B - Condensed Matter and Materials Physics}\
  }\textbf {\bibinfo {volume} {86}},\ \bibinfo {pages} {1} (\bibinfo {year}
  {2012})}\BibitemShut {NoStop}%
\bibitem [{\citenamefont {Alsubaie}\ \emph {et~al.}(2017)\citenamefont
  {Alsubaie}, \citenamefont {Sharma}, \citenamefont {Liu}, \citenamefont
  {Nagarajan},\ and\ \citenamefont {Seidel}}]{Alsubaie2017}%
  \BibitemOpen
  \bibfield  {author} {\bibinfo {author} {\bibfnamefont {A.}~\bibnamefont
  {Alsubaie}}, \bibinfo {author} {\bibfnamefont {P.}~\bibnamefont {Sharma}},
  \bibinfo {author} {\bibfnamefont {G.}~\bibnamefont {Liu}}, \bibinfo {author}
  {\bibfnamefont {V.}~\bibnamefont {Nagarajan}},\ and\ \bibinfo {author}
  {\bibfnamefont {J.}~\bibnamefont {Seidel}},\ }\bibfield  {title} {\bibinfo
  {title} {{Mechanical stress-induced switching kinetics of ferroelectric thin
  films at the nanoscale}},\ }\bibfield  {journal} {\bibinfo  {journal}
  {Nanotechnology}\ }\textbf {\bibinfo {volume} {28}},\ \href
  {https://doi.org/10.1088/1361-6528/aa536d} {10.1088/1361-6528/aa536d}
  (\bibinfo {year} {2017})\BibitemShut {NoStop}%
\bibitem [{\citenamefont {Ederer}\ and\ \citenamefont
  {Spaldin}(2005)}]{Ederer2005}%
  \BibitemOpen
  \bibfield  {author} {\bibinfo {author} {\bibfnamefont {C.}~\bibnamefont
  {Ederer}}\ and\ \bibinfo {author} {\bibfnamefont {N.~A.}\ \bibnamefont
  {Spaldin}},\ }\bibfield  {title} {\bibinfo {title} {{Effect of epitaxial
  strain on the spontaneous polarization of thin film ferroelectrics}},\ }\href
  {https://doi.org/10.1103/PhysRevLett.95.257601} {\bibfield  {journal}
  {\bibinfo  {journal} {Physical Review Letters}\ }\textbf {\bibinfo {volume}
  {95}},\ \bibinfo {pages} {2} (\bibinfo {year} {2005})},\ \Eprint
  {https://arxiv.org/abs/0508005} {arXiv:0508005 [cond-mat]} \BibitemShut
  {NoStop}%
\bibitem [{\citenamefont {Chen}\ \emph {et~al.}(2018)\citenamefont {Chen},
  \citenamefont {Bai}, \citenamefont {Zhang},\ and\ \citenamefont
  {Jiang}}]{Chen2018}%
  \BibitemOpen
  \bibfield  {author} {\bibinfo {author} {\bibfnamefont {D.}~\bibnamefont
  {Chen}}, \bibinfo {author} {\bibfnamefont {Z.}~\bibnamefont {Bai}}, \bibinfo
  {author} {\bibfnamefont {Y.}~\bibnamefont {Zhang}},\ and\ \bibinfo {author}
  {\bibfnamefont {A.}~\bibnamefont {Jiang}},\ }\bibfield  {title} {\bibinfo
  {title} {{Strain induced enhancement of erasable domain wall current in
  epitaxial BiFeO 3 thin films}},\ }\href {https://doi.org/10.1063/1.5054945}
  {\bibfield  {journal} {\bibinfo  {journal} {Journal of Applied Physics}\
  }\textbf {\bibinfo {volume} {124}},\ \bibinfo {pages} {194102} (\bibinfo
  {year} {2018})}\BibitemShut {NoStop}%
\bibitem [{\citenamefont {Schlom}\ \emph {et~al.}(2007)\citenamefont {Schlom},
  \citenamefont {Chen}, \citenamefont {Eom}, \citenamefont {Rabe},
  \citenamefont {Streiffer},\ and\ \citenamefont {Triscone}}]{Schlom2007}%
  \BibitemOpen
  \bibfield  {author} {\bibinfo {author} {\bibfnamefont {D.~G.}\ \bibnamefont
  {Schlom}}, \bibinfo {author} {\bibfnamefont {L.-Q.}\ \bibnamefont {Chen}},
  \bibinfo {author} {\bibfnamefont {C.-B.}\ \bibnamefont {Eom}}, \bibinfo
  {author} {\bibfnamefont {K.~M.}\ \bibnamefont {Rabe}}, \bibinfo {author}
  {\bibfnamefont {S.~K.}\ \bibnamefont {Streiffer}},\ and\ \bibinfo {author}
  {\bibfnamefont {J.-M.}\ \bibnamefont {Triscone}},\ }\bibfield  {title}
  {\bibinfo {title} {{Strain Tuning of Ferroelectric Thin Films}},\ }\href
  {https://doi.org/10.1146/annurev.matsci.37.061206.113016} {\bibfield
  {journal} {\bibinfo  {journal} {Annual Review of Materials Research}\
  }\textbf {\bibinfo {volume} {37}},\ \bibinfo {pages} {589} (\bibinfo {year}
  {2007})}\BibitemShut {NoStop}%
\bibitem [{\citenamefont {Kirbus}\ \emph
  {et~al.}(2019{\natexlab{b}})\citenamefont {Kirbus}, \citenamefont {Godau},
  \citenamefont {Wehmeier}, \citenamefont {Beccard}, \citenamefont
  {Beyreuther}, \citenamefont {Hau{\ss}mann},\ and\ \citenamefont
  {Eng}}]{Kirbus2019a}%
  \BibitemOpen
  \bibfield  {author} {\bibinfo {author} {\bibfnamefont {B.}~\bibnamefont
  {Kirbus}}, \bibinfo {author} {\bibfnamefont {C.}~\bibnamefont {Godau}},
  \bibinfo {author} {\bibfnamefont {L.}~\bibnamefont {Wehmeier}}, \bibinfo
  {author} {\bibfnamefont {H.}~\bibnamefont {Beccard}}, \bibinfo {author}
  {\bibfnamefont {E.}~\bibnamefont {Beyreuther}}, \bibinfo {author}
  {\bibfnamefont {A.}~\bibnamefont {Hau{\ss}mann}},\ and\ \bibinfo {author}
  {\bibfnamefont {L.~M.}\ \bibnamefont {Eng}},\ }\bibfield  {title} {\bibinfo
  {title} {{Real-Time 3D Imaging of Nanoscale Ferroelectric Domain Wall
  Dynamics in Lithium Niobate Single Crystals under Electric Stimuli:
  Implications for Domain-Wall-Based Nanoelectronic Devices}},\ }\href
  {https://doi.org/10.1021/acsanm.9b01240} {\bibfield  {journal} {\bibinfo
  {journal} {ACS Applied Nano Materials}\ }\textbf {\bibinfo {volume} {2}},\
  \bibinfo {pages} {5787} (\bibinfo {year} {2019}{\natexlab{b}})}\BibitemShut
  {NoStop}%
\bibitem [{\citenamefont {Hicks}\ \emph {et~al.}(2014)\citenamefont {Hicks},
  \citenamefont {Barber}, \citenamefont {Edkins}, \citenamefont {Brodsky},\
  and\ \citenamefont {Mackenzie}}]{Hicks2014}%
  \BibitemOpen
  \bibfield  {author} {\bibinfo {author} {\bibfnamefont {C.~W.}\ \bibnamefont
  {Hicks}}, \bibinfo {author} {\bibfnamefont {M.~E.}\ \bibnamefont {Barber}},
  \bibinfo {author} {\bibfnamefont {S.~D.}\ \bibnamefont {Edkins}}, \bibinfo
  {author} {\bibfnamefont {D.~O.}\ \bibnamefont {Brodsky}},\ and\ \bibinfo
  {author} {\bibfnamefont {A.~P.}\ \bibnamefont {Mackenzie}},\ }\bibfield
  {title} {\bibinfo {title} {{Piezoelectric-based apparatus for strain
  tuning}},\ }\href {https://doi.org/10.1063/1.4881611} {\bibfield  {journal}
  {\bibinfo  {journal} {Review of Scientific Instruments}\ }\textbf {\bibinfo
  {volume} {85}},\ \bibinfo {pages} {065003} (\bibinfo {year}
  {2014})}\BibitemShut {NoStop}%
\bibitem [{\citenamefont {Yamada}\ \emph {et~al.}(1967)\citenamefont {Yamada},
  \citenamefont {Niizeki},\ and\ \citenamefont {Toyoda}}]{Yamada1967}%
  \BibitemOpen
  \bibfield  {author} {\bibinfo {author} {\bibfnamefont {T.}~\bibnamefont
  {Yamada}}, \bibinfo {author} {\bibfnamefont {N.}~\bibnamefont {Niizeki}},\
  and\ \bibinfo {author} {\bibfnamefont {H.}~\bibnamefont {Toyoda}},\
  }\bibfield  {title} {\bibinfo {title} {{Piezoelectric and Elastic Properties
  of Lithium Niobate Single Crystals}},\ }\href
  {https://doi.org/10.1143/JJAP.6.151} {\bibfield  {journal} {\bibinfo
  {journal} {Japanese Journal of Applied Physics}\ }\textbf {\bibinfo {volume}
  {6}},\ \bibinfo {pages} {151} (\bibinfo {year} {1967})}\BibitemShut {NoStop}%
\bibitem [{\citenamefont {Smith}\ and\ \citenamefont
  {Welsh}(1971)}]{Smith1971a}%
  \BibitemOpen
  \bibfield  {author} {\bibinfo {author} {\bibfnamefont {R.~T.}\ \bibnamefont
  {Smith}}\ and\ \bibinfo {author} {\bibfnamefont {F.~S.}\ \bibnamefont
  {Welsh}},\ }\bibfield  {title} {\bibinfo {title} {{Temperature Dependence of
  the Elastic, Piezoelectric, and Dielectric Constants of Lithium Tantalate and
  Lithium Niobate}},\ }\href {https://doi.org/10.1063/1.1660528} {\bibfield
  {journal} {\bibinfo  {journal} {Journal of Applied Physics}\ }\textbf
  {\bibinfo {volume} {42}},\ \bibinfo {pages} {2219} (\bibinfo {year}
  {1971})}\BibitemShut {NoStop}%
\bibitem [{\citenamefont {Weis}\ and\ \citenamefont
  {Gaylord}(1985)}]{Weis1985}%
  \BibitemOpen
  \bibfield  {author} {\bibinfo {author} {\bibfnamefont {R.~S.}\ \bibnamefont
  {Weis}}\ and\ \bibinfo {author} {\bibfnamefont {T.~K.}\ \bibnamefont
  {Gaylord}},\ }\bibfield  {title} {\bibinfo {title} {{Lithium niobate: Summary
  of physical properties and crystal structure}},\ }\href
  {https://doi.org/10.1007/BF00614817} {\bibfield  {journal} {\bibinfo
  {journal} {Applied Physics A Solids and Surfaces}\ }\textbf {\bibinfo
  {volume} {37}},\ \bibinfo {pages} {191} (\bibinfo {year} {1985})}\BibitemShut
  {NoStop}%
\bibitem [{\citenamefont {Wemple}\ \emph {et~al.}(1968)\citenamefont {Wemple},
  \citenamefont {DiDomenico},\ and\ \citenamefont {Camlibel}}]{Wemple1968a}%
  \BibitemOpen
  \bibfield  {author} {\bibinfo {author} {\bibfnamefont {S.~H.}\ \bibnamefont
  {Wemple}}, \bibinfo {author} {\bibfnamefont {M.}~\bibnamefont {DiDomenico}},\
  and\ \bibinfo {author} {\bibfnamefont {I.}~\bibnamefont {Camlibel}},\
  }\bibfield  {title} {\bibinfo {title} {{ Relationship between linear and
  quadratic electrooptic coefficients in LiNbO3, LiTiO3, and other oxygen
  octahedra ferroelectrics based on direct measurement of spontaneous
  polarization}},\ }\href {https://doi.org/10.1063/1.1651955} {\bibfield
  {journal} {\bibinfo  {journal} {Applied Physics Letters}\ }\textbf {\bibinfo
  {volume} {12}},\ \bibinfo {pages} {209} (\bibinfo {year} {1968})}\BibitemShut
  {NoStop}%
\bibitem [{\citenamefont {Xiao}\ \emph {et~al.}(2018)\citenamefont {Xiao},
  \citenamefont {K{\"{a}}mpfe}, \citenamefont {Jin}, \citenamefont
  {Hau{\ss}mann}, \citenamefont {Lu},\ and\ \citenamefont {Eng}}]{Xiao2018}%
  \BibitemOpen
  \bibfield  {author} {\bibinfo {author} {\bibfnamefont {S.~Y.}\ \bibnamefont
  {Xiao}}, \bibinfo {author} {\bibfnamefont {T.}~\bibnamefont {K{\"{a}}mpfe}},
  \bibinfo {author} {\bibfnamefont {Y.~M.}\ \bibnamefont {Jin}}, \bibinfo
  {author} {\bibfnamefont {A.}~\bibnamefont {Hau{\ss}mann}}, \bibinfo {author}
  {\bibfnamefont {X.~M.}\ \bibnamefont {Lu}},\ and\ \bibinfo {author}
  {\bibfnamefont {L.~M.}\ \bibnamefont {Eng}},\ }\bibfield  {title} {\bibinfo
  {title} {{Dipole-Tunneling Model from Asymmetric Domain-Wall Conductivity in
  LiNb O3 Single Crystals}},\ }\href
  {https://doi.org/10.1103/PhysRevApplied.10.034002} {\bibfield  {journal}
  {\bibinfo  {journal} {Physical Review Applied}\ }\textbf {\bibinfo {volume}
  {10}},\ \bibinfo {pages} {1} (\bibinfo {year} {2018})}\BibitemShut {NoStop}%
\bibitem [{\citenamefont {Wolba}\ \emph {et~al.}(2018)\citenamefont {Wolba},
  \citenamefont {Seidel}, \citenamefont {Cazorla}, \citenamefont {Godau},
  \citenamefont {Hau{\ss}mann},\ and\ \citenamefont {Eng}}]{Wolba2018}%
  \BibitemOpen
  \bibfield  {author} {\bibinfo {author} {\bibfnamefont {B.}~\bibnamefont
  {Wolba}}, \bibinfo {author} {\bibfnamefont {J.}~\bibnamefont {Seidel}},
  \bibinfo {author} {\bibfnamefont {C.}~\bibnamefont {Cazorla}}, \bibinfo
  {author} {\bibfnamefont {C.}~\bibnamefont {Godau}}, \bibinfo {author}
  {\bibfnamefont {A.}~\bibnamefont {Hau{\ss}mann}},\ and\ \bibinfo {author}
  {\bibfnamefont {L.~M.}\ \bibnamefont {Eng}},\ }\bibfield  {title} {\bibinfo
  {title} {{Resistor Network Modeling of Conductive Domain Walls in Lithium
  Niobate}},\ }\href {https://doi.org/10.1002/aelm.201700242} {\bibfield
  {journal} {\bibinfo  {journal} {Advanced Electronic Materials}\ }\textbf
  {\bibinfo {volume} {4}},\ \bibinfo {pages} {1700242} (\bibinfo {year}
  {2018})}\BibitemShut {NoStop}%
\bibitem [{\citenamefont {M{\"{u}}ller}\ \emph {et~al.}(2003)\citenamefont
  {M{\"{u}}ller}, \citenamefont {Soergel},\ and\ \citenamefont
  {Buse}}]{Muller2003}%
  \BibitemOpen
  \bibfield  {author} {\bibinfo {author} {\bibfnamefont {M.}~\bibnamefont
  {M{\"{u}}ller}}, \bibinfo {author} {\bibfnamefont {E.}~\bibnamefont
  {Soergel}},\ and\ \bibinfo {author} {\bibfnamefont {K.}~\bibnamefont
  {Buse}},\ }\bibfield  {title} {\bibinfo {title} {{Influence of ultraviolet
  illumination on the poling characteristics of lithium niobate crystals}},\
  }\href {https://doi.org/10.1063/1.1606504} {\bibfield  {journal} {\bibinfo
  {journal} {Applied Physics Letters}\ }\textbf {\bibinfo {volume} {83}},\
  \bibinfo {pages} {1824} (\bibinfo {year} {2003})}\BibitemShut {NoStop}%
\bibitem [{\citenamefont {K{\"{a}}mpfe}\ \emph {et~al.}(2014)\citenamefont
  {K{\"{a}}mpfe}, \citenamefont {Reichenbach}, \citenamefont {Schr{\"{o}}der},
  \citenamefont {Hau{\ss}mann}, \citenamefont {Eng}, \citenamefont {Woike},\
  and\ \citenamefont {Soergel}}]{Kampfe2014}%
  \BibitemOpen
  \bibfield  {author} {\bibinfo {author} {\bibfnamefont {T.}~\bibnamefont
  {K{\"{a}}mpfe}}, \bibinfo {author} {\bibfnamefont {P.}~\bibnamefont
  {Reichenbach}}, \bibinfo {author} {\bibfnamefont {M.}~\bibnamefont
  {Schr{\"{o}}der}}, \bibinfo {author} {\bibfnamefont {A.}~\bibnamefont
  {Hau{\ss}mann}}, \bibinfo {author} {\bibfnamefont {L.~M.}\ \bibnamefont
  {Eng}}, \bibinfo {author} {\bibfnamefont {T.}~\bibnamefont {Woike}},\ and\
  \bibinfo {author} {\bibfnamefont {E.}~\bibnamefont {Soergel}},\ }\bibfield
  {title} {\bibinfo {title} {{Optical three-dimensional profiling of charged
  domain walls in ferroelectrics by Cherenkov second-harmonic generation}},\
  }\href {https://doi.org/10.1103/PhysRevB.89.035314} {\bibfield  {journal}
  {\bibinfo  {journal} {Physical Review B}\ }\textbf {\bibinfo {volume} {89}},\
  \bibinfo {pages} {035314} (\bibinfo {year} {2014})}\BibitemShut {NoStop}%
\bibitem [{\citenamefont {Soergel}(2011)}]{Soergel2011}%
  \BibitemOpen
  \bibfield  {author} {\bibinfo {author} {\bibfnamefont {E.}~\bibnamefont
  {Soergel}},\ }\bibfield  {title} {\bibinfo {title} {{Piezoresponse force
  microscopy (PFM)}},\ }\href {https://doi.org/10.1088/0022-3727/44/46/464003}
  {\bibfield  {journal} {\bibinfo  {journal} {Journal of Physics D: Applied
  Physics}\ }\textbf {\bibinfo {volume} {44}},\ \bibinfo {pages} {464003}
  (\bibinfo {year} {2011})}\BibitemShut {NoStop}%
\bibitem [{\citenamefont {Eliseev}\ \emph {et~al.}(2011)\citenamefont
  {Eliseev}, \citenamefont {Morozovska}, \citenamefont {Svechnikov},
  \citenamefont {Gopalan},\ and\ \citenamefont {Shur}}]{Eliseev2011}%
  \BibitemOpen
  \bibfield  {author} {\bibinfo {author} {\bibfnamefont {E.~A.}\ \bibnamefont
  {Eliseev}}, \bibinfo {author} {\bibfnamefont {A.~N.}\ \bibnamefont
  {Morozovska}}, \bibinfo {author} {\bibfnamefont {G.~S.}\ \bibnamefont
  {Svechnikov}}, \bibinfo {author} {\bibfnamefont {V.}~\bibnamefont
  {Gopalan}},\ and\ \bibinfo {author} {\bibfnamefont {V.~Y.}\ \bibnamefont
  {Shur}},\ }\bibfield  {title} {\bibinfo {title} {{Static conductivity of
  charged domain walls in uniaxial ferroelectric semiconductors}},\ }\href
  {https://doi.org/10.1103/PhysRevB.83.235313} {\bibfield  {journal} {\bibinfo
  {journal} {Physical Review B - Condensed Matter and Materials Physics}\
  }\textbf {\bibinfo {volume} {83}},\ \bibinfo {pages} {1} (\bibinfo {year}
  {2011})}\BibitemShut {NoStop}%
\end{thebibliography}%

\end{document}



\title{Supplementary material to: Tuning the domain wall conductivity in bulk lithium niobate by uniaxial stress}

\author{Ekta Singh}
\affiliation{Institute of Applied Physics, TU Dresden, N{\"o}thnitzer Stra{\ss}e 61, 01187 Dresden, Germany }
\affiliation{Max Planck Institute for Chemical Physics of Solids, N{\"o}thnitzer Stra{\ss}e 40, 01187 Dresden, Germany}
\author{Henrik Beccard}

\affiliation{Institute of Applied Physics, TU Dresden, N{\"o}thnitzer Stra{\ss}e 61, 01187 Dresden, Germany }

\author{Zeeshan H. Amber}

\affiliation{Institute of Applied Physics, TU Dresden, N{\"o}thnitzer Stra{\ss}e 61, 01187 Dresden, Germany }

\author{Julius Ratzenberger}
\affiliation{Institute of Applied Physics, TU Dresden, N{\"o}thnitzer Stra{\ss}e 61, 01187 Dresden, Germany }

\author{Clifford W. Hicks}
\affiliation{Max Planck Institute for Chemical Physics of Solids, N{\"o}thnitzer Stra{\ss}e 40, 01187 Dresden, Germany} 
\affiliation{School of Physics and Astronomy, University of Birmingham, Birmingham B15 2TT, United Kingdom}

\author{Michael R{\"u}sing}

\affiliation{Institute of Applied Physics, TU Dresden, N{\"o}thnitzer Stra{\ss}e 61, 01187 Dresden, Germany }

\author{Lukas M. Eng}

\affiliation{Institute of Applied Physics, TU Dresden, N{\"o}thnitzer Stra{\ss}e 61, 01187 Dresden, Germany }
 
\affiliation{ ct.qmat: Dresden-W{\"u}rzburg Cluster of Excellence-EXC 2147, TU Dresden, 01062 Dresden, Germany}
\date{\today}

\maketitle
\section{S1. supplementary Methods}
\subsection{S1.1. Macroscopic mesurements}
\begin{figure}[h!]
\includegraphics{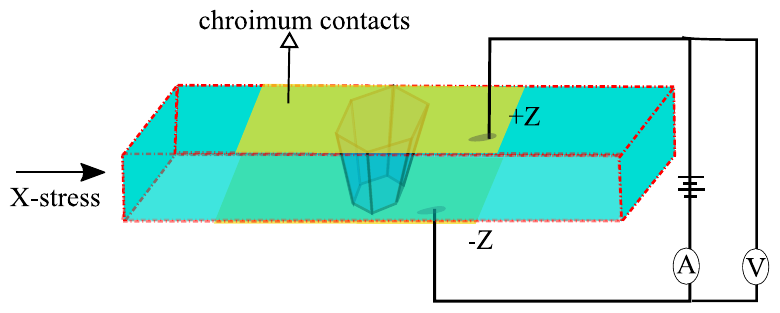}
\caption{\label{fig:setup} Setup for macroscopic measurement. The domain wall in the sample is connected to Keithley by chromium contacts on both sides.}
\end{figure}

\begin{figure}[h!]
\includegraphics[width=5in]{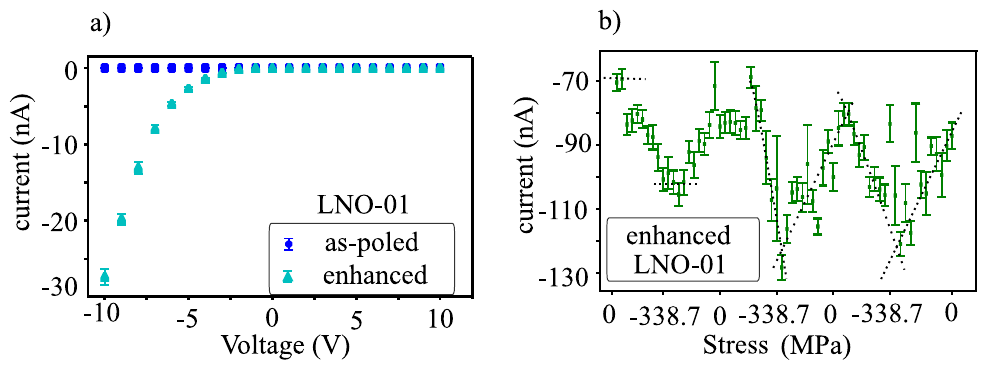}
\caption{\label{fig:s1} I-V characteristics of a)  as-poled and enhanced domain walls of sample LNO-01 (thickness: 200 $\upmu$m). b) macroscopic response to uniaxial stress along x-axis at -20 V of sample LNO-01.}
\end{figure}

Here, we show the I-V curves and macroscopic behaviour of the LNO-01 sample with stress, which is discussed in the main paper. Here, instead of locally measuring the current by cAFM the global response is measured. In Fig.~\ref{fig:s1}(a) we have plotted I-V curves acquired with macroscopic chromium electrodes using a Keithley 6517B electrometer. Fig.~\ref{fig:s1}(a) shows I-V curves  before and after enhancement, where the current has increased by the factor of 10$^4$ in enhanced domain walls. The diode and ohmic nature of the I-V curve varies from sample to sample as it depends on the barrier at the interface of sample and electrodes. In fig.~\ref{fig:s1}(b) we measured the effect of applied stress on the conductivity of domain walls via macroscopic electrodes at the voltage of -20 V, because we see no current in direction of positive voltage. The sample shows a positive relative current change of 0.48 $\pm$ 0.02 with respect to applied stress in first cycle which indicates that induced Head-to-Head i(h2h) domain walls  are dominating the behaviour.

\begin{figure}[h!]
\includegraphics[width=5in]{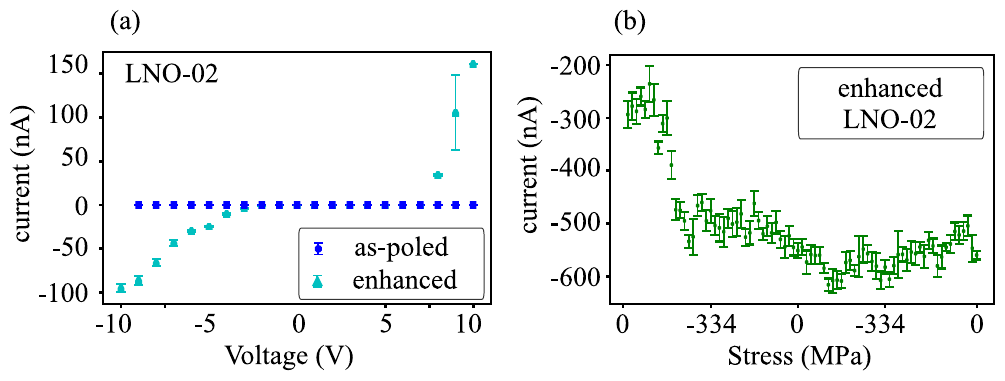}
\caption{\label{fig:s2} I-V characteristics of a)  as-poled and enhanced domain walls of sample LNO-02 (thickness: 200 $\upmu$m). b) macroscopic response to uniaxial stress along x-axis at -20 V of sample LNO-02.}
\end{figure}

Sample LNO-02 shows a similar increase in domain wall current with enhancement, see Fig.\ref{fig:s2}(a) . When stress was applied to the sample LNO-02 and current was measured with same electric field as was applied in sample LNO-01, we saw that the response of current flowing through domain walls is non-conclusive, see in Fig.~\ref{fig:s2}(b) i.e. no global trend is observed. As discussed in the main paper, this macroscopic behavior is a result of participation of all induced types domain walls. This participation could be mainly of induced head-to-head or induced tail-to-tail or even both types of walls. Overall, we conclude that stress can change the domain wall conductivity. For a detailed, microscopic interpretation of the observed behaviour cAFM measurements were shown in the main paper and also in this supplement.

\subsection{S1.2. Extraction of domain wall inclination angles}
\begin{figure}[h!]
\includegraphics{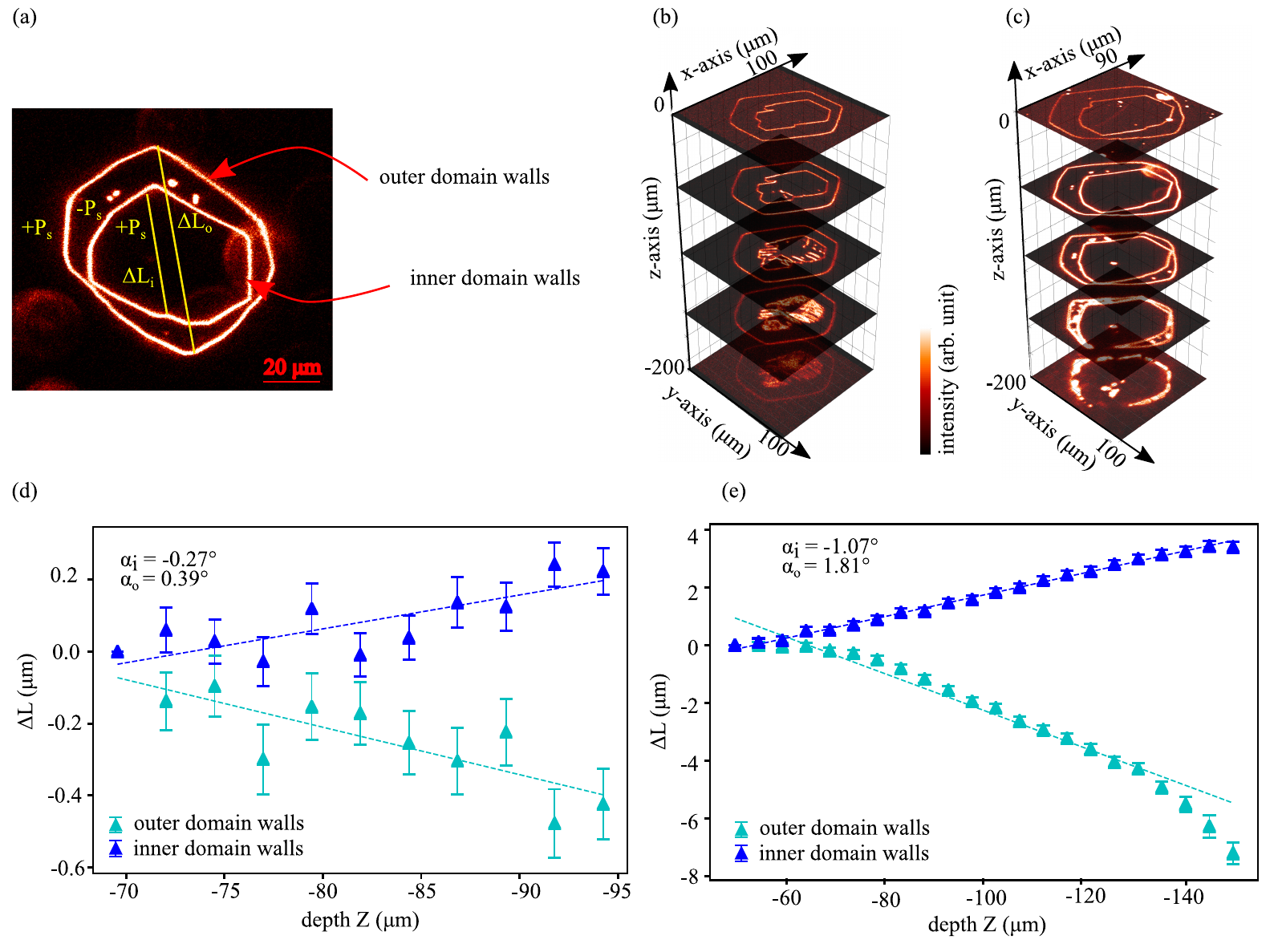}
\caption{\label{fig:angle_cal}(a) Surface second harmonic generation (SHG) image of enhanced domain walls in sample LNO-01, where lines going through the domain walls show the location where angle is calculated in both subfigures (d) and (e) of this image.(b) and (c) SHG scans at different depths of as-poled and enhanced DWs in sample LNO-01, respectively. (d) Angle calculation for the inner and the outer as-poled domain walls, where x-axis is depth of the sample and y-axis is width of the domain wall.  (e) Angle calculation for the outer enhanced domain walls, where x-axis is depth of the sample and y-axis is width of the domain wall.  }
\end{figure}
\begin{figure}[h!]
\includegraphics{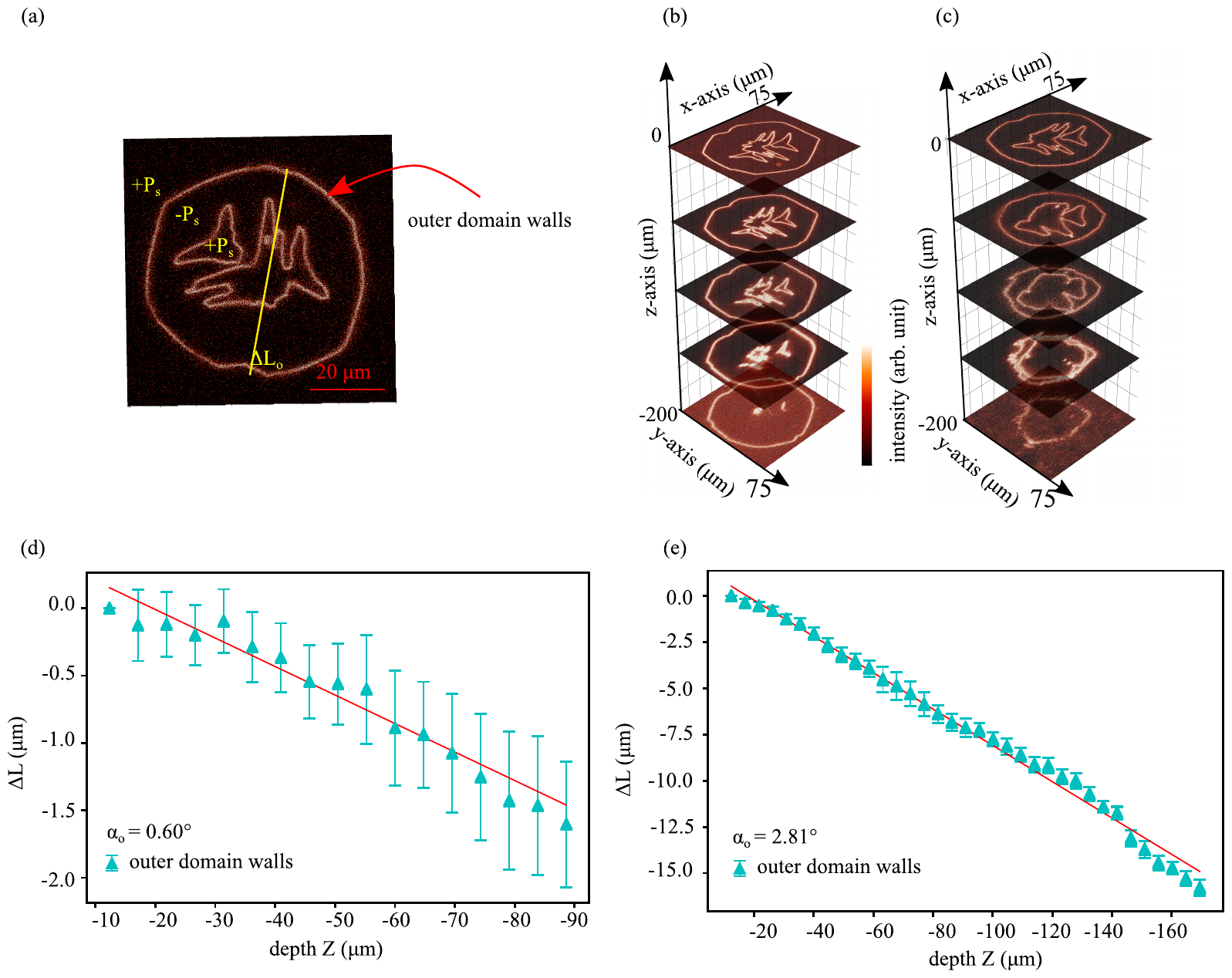}
\caption{\label{fig:angle_cal_1}  (a) surface second harmonic generation (SHG) image of domain walls in sample LNO-02, where a line going through the domain wall shows the location where an angle is calculated. (b) and (c) SHG scans at different depths of as-poled and enhanced DWs in sample LNO-02, respectively. (d) Angle calculation for the outer as-poled domain walls, where x-axis is depth of the sample and y-axis is width of the domain wall.  (e) Angle calculation for the outer enhanced domain walls, where x-axis is depth of the sample and y-axis is width of the domain wall.  }
\end{figure}
In Fig.~\ref{fig:angle_cal} we show the calculation of the average angle of the domain walls with the help of 3D second harmonic generation data. Widths of the both inner and outer domain walls at different depths were taken and were plotted as a function of the depth based on 3D SHG scan. Imaging was performed at wavelength of 900 nm and the backscattered SHG signal was detected at 450 nm wavelength. An air objective with magnification of 20X and numerical aperture (NA) of 0.8 was used. The pixel sizes were kept as following: for as-poled domain walls, pixel size was 0.414 $\times$ 0.414 $\upmu$m$^2$ and for enhanced domain walls 0.207 $\times$ 0.207 $\upmu$m$^2$. To obtain the location of a domain wall, we fitted the intensity profile of the SHG signal at the domain wall with a Gaussian function and chose its maxima as the domain wall location. The calculated angle for (1) the inner domain wall is $\alpha_i = -0.27^\circ$ $\pm$ 0.08$^\circ$, (2) the outer domain wall is $\alpha_o = 0.39^\circ$ $\pm$ 0.10$^\circ$.  For the enhanced LNO-01 sample the calculated angle for (1) the inner domain wall is $\alpha_i = -1.07^\circ$ $\pm$ 0.03$^\circ$, (2) the outer domain wall is $\alpha_o = 1.81^\circ$ $\pm$ 0.09$^\circ$. These values can be interpreted as an average angles at the lines passing through a corner of domain walls in Fig.~\ref{fig:angle_cal}(a).

A similar procedure was used to calculate the domain wall inclination angle ($\alpha$) for sample LNO-02. In Fig.~\ref{fig:angle_cal_1}(a), we show the SHG image of the top surface of sample LNO-02. The sample consists of two domains and we have calculated the $\alpha$ only for outer domain wall because the inner domain wall has no regular shape. Similar to the sample LNO-01, sample LNO-02 also shows an increase in the the inclination angle for enhanced domain walls. Outer domain wall in sample LNO-02 has higher $\alpha$ than in the sample LNO-01. As can be seen from Fig.~\ref{fig:angle_cal_1}(d) and (e) where $\alpha$ increased from 0.60$^{\circ}$ $\pm$ 0.04$^\circ$ to 2.81$^{\circ}$ $\pm$ 0.04$^\circ$ from as-poled to enhanced domain walls respectively.

\newpage
\section{S2. supplementary calculations}
\subsection{S2.1. Calculation of the induced polarization via direct piezoelectric effect}

In this section we provide the detailed calculation of the induced polarization. Here, we have divided the calculations in two different cases for different stress axes. 

Case 1: No external field is applied to the sample. 

Case 2: An external field of 100 kV/m is applied along z-axis.\\

Starting with the general expression of the direct piezoelectricity.
\begin{equation}
[D] = [d][X]+[\epsilon^X][E] \label{eq:polarization1}
\end{equation}	

where [D] is the matrix of displacement field vector, [d] is a rank-3 piezoelectric strain coefficient tensor, [X] is a rank-2 stress tensor, [$\epsilon^X$] is the dielectric permittivity at constant stress, and [E] is the electric field. 

There, the electric displacement field is also related to the polarization by:
\begin{equation}\label{eq:polarization2}
[D]=[P]+\epsilon_o [E]
\end{equation}

\textbf{ CASE 1(x-stress):}  When no electric field is applied to the sample

From equation \ref{eq:polarization1} and \ref{eq:polarization2} follows: 

\begin{gather}
[P]=[d][X]	\\ \label{eq:stress and polarization}
P_{j} = d_{jkl}X_{kl}\\ 
d_{jkl} \neq d_{klj}
\end{gather}

\par LiNbO$_3$ belongs to a trigonal crystal system and its ferroelectric phase belongs to the 3m point group. So, the tensors for point group 3m are used in the equation \ref{eq:stress and polarization}:
\begin{gather}
	\begin{bmatrix}
		P_1 \\P_2 \\P_3 
	\end{bmatrix}
	= 
	\begin{bmatrix}
		0&0&0&0&d_{15}&-2d_{22}\\-d_{22}&d_{22}&0&d_{15}&0&0\\d_{31}&d_{31}&d_{33}&0&0&0
	\end{bmatrix}
	\begin{bmatrix}
		X_1\\X_2\\X_3\\X_4\\X_5\\X_6
	\end{bmatrix}
\end{gather}
For uniaxial stress along the x-axis we have, $X_1 \neq 0 ;  X_2=X_3=X_4=X_5=X_6=0$, where $kl$ notations in equation 4 are: $ 11=1, 22=2, 33=3, 32=23=4, 13=31=5, 21=12=6$. For a compression of, $X_1$ = -100 MPa, the values of induced polarization at the room temperature using $\ d_{22}= 20.8 \times 10^{-12}$ C/N, $\ d_{31}=-0.863 \times 10^{-12}$ C/N can be calculated as:
\begin{gather}
P_1 = d_{15}X_5-2d_{22}X_6 = \textbf{0}\\
P_2 = -d_{22}X_1+d_{22}X_2+d_{15}X_4 = -d_{22}X_1 = \textbf{0.208 $\times$ 10$^{-2}$ Cm$^{-2}$ }\\
P_3 = d_{31}X_1	+d_{31}X_2+d_{33}X_3 = d_{31}X_1 = \textbf{0.863 $\times$ 10$^{-4}$ Cm$^{-2}$}
\end{gather}

\textbf{CASE 2 (x-stress):} In case of the external electric field applied to sample along z-axis.\\
We get:

\begin{gather}
	[P]=[d][X] +\epsilon_o [E]-[\epsilon^X][E]\\
	[P]=[d][X]+([\epsilon^X]-\epsilon_o)[E]\\
	\begin{bmatrix} \label{eq:with electric field}
		P_1 \\P_2 \\P_3 
	\end{bmatrix}
	= 
	\begin{bmatrix}
		0&0&0&0&d_{15}&-2d_{22}\\-d_{22}&d_{22}&0&d_{15}&0&0\\d_{31}&d_{31}&d_{33}&0&0&0
	\end{bmatrix}
	\begin{bmatrix}
		X_1\\X_2\\X_3\\X_4\\X_5\\X_6
	\end{bmatrix} + 
	\begin{bmatrix}
	\epsilon_{11}^X - \epsilon_o & -\epsilon_o & -\epsilon_o \\
	-\epsilon_o &\epsilon_{22}^X - \epsilon_o & -\epsilon_o \\
	-\epsilon_o & -\epsilon_o & \epsilon_{33}^X- \epsilon_o
	\end{bmatrix}
	\begin{bmatrix}
		E_1\\
		E_2\\
		E_3
	\end{bmatrix}
\end{gather}

Here, E$_1$ =E$_2$ = 0 and E$_3$ = 100 kVm$^{-1}$. In our case as explained above, using  equation \ref{eq:with electric field} and the values $\epsilon_{11}^X$=$\epsilon_{22}^X$ = 85.2 $\epsilon_o$, $\epsilon_{33}^X$ = 28.7 $\epsilon_o$, $\epsilon_o$ = 8.85 $\times$ 10$^{-12}$ C V$^{-1}$m$^{-1}$, one can calculate induced polarization for x-compression of -100 MPa:  
\begin{gather}
P_{1} = -\epsilon_o E_3 = \textbf{-0.885 $\times$ 10$^{-6}$ C m$^{-2}$}\\
P_{2} = -d_{22} X_1 -\epsilon_o E_3 = \textbf{0.2079 $\times$ 10$^{-2}$ Cm$^{-2}$}\\	 
P_{3} = d_{31} X_1 - (\epsilon_{33}^X -\epsilon_o)E_3 = \textbf{0.618 $\times$ 10$^{-4}$ Cm$^{-2}$}
\end{gather}
\\
$-------$\\
For uniaxial compression along y-axis we have, $X_2$ = -100  MPa. All other stress components become zero and then following the same procedure one can calculate for :\\
\textbf{ CASE 1 (y-stress):}  when no electric field is applied to sample\\
\begin{gather}
P_1 = d_{15}X_5-2d_{22}X_6 = \textbf{0}\\
P_2 = -d_{22}X_1+d_{22}X_2+d_{15}X_4 = d_{22}X_2 = \textbf{-0.208 $\times$ 10$^{-2}$ Cm$^{-2}$ }\\
P_3 = d_{31}X_1	+d_{31}X_2+d_{33}X_3 = d_{31}X_2 = \textbf{0.863 $\times$ 10$^{-4}$ Cm$^{-2}$}
\end{gather}

\textbf{CASE 2 (y-stress):} In case of the external electric field applied to sample along z-axis\\
\begin{gather}
P_{1} = -\epsilon_o E_3 = \textbf{-0.885 $\times$ 10$^{-6}$ C m$^{-2}$}\\
P_{2} = d_{22} X_2 -\epsilon_o E_3 = \textbf{-0.208 $\times$ 10$^{-2}$ Cm$^{-2}$}\\	 
P_{3} = d_{31} X_2 - (\epsilon_{33}^X -\epsilon_o)E_3 = \textbf{0.618 $\times$ 10$^{-4}$ Cm$^{-2}$}
\end{gather}
As can be seen from the numerical values, the change in the induced polarization, when an external electric field is applied, is in fourth digit after decimal, which can be neglected compared to the piezoelectric contribution. This is the reason why we did not consider the contribution of the electric field in the main paper.

The other very important message to take from this calculation is that the sign of $\Delta P_y$ changes when sample is compressed along Y-axis(LNO-02) compared to when sample is compressed along x-axis(LNO-01).

\newpage
\section{S3. supplementary results}
\subsection{S3.1. Microscopic cAFM results for intermediate stress steps}

\begin{figure*}[!h]
\includegraphics{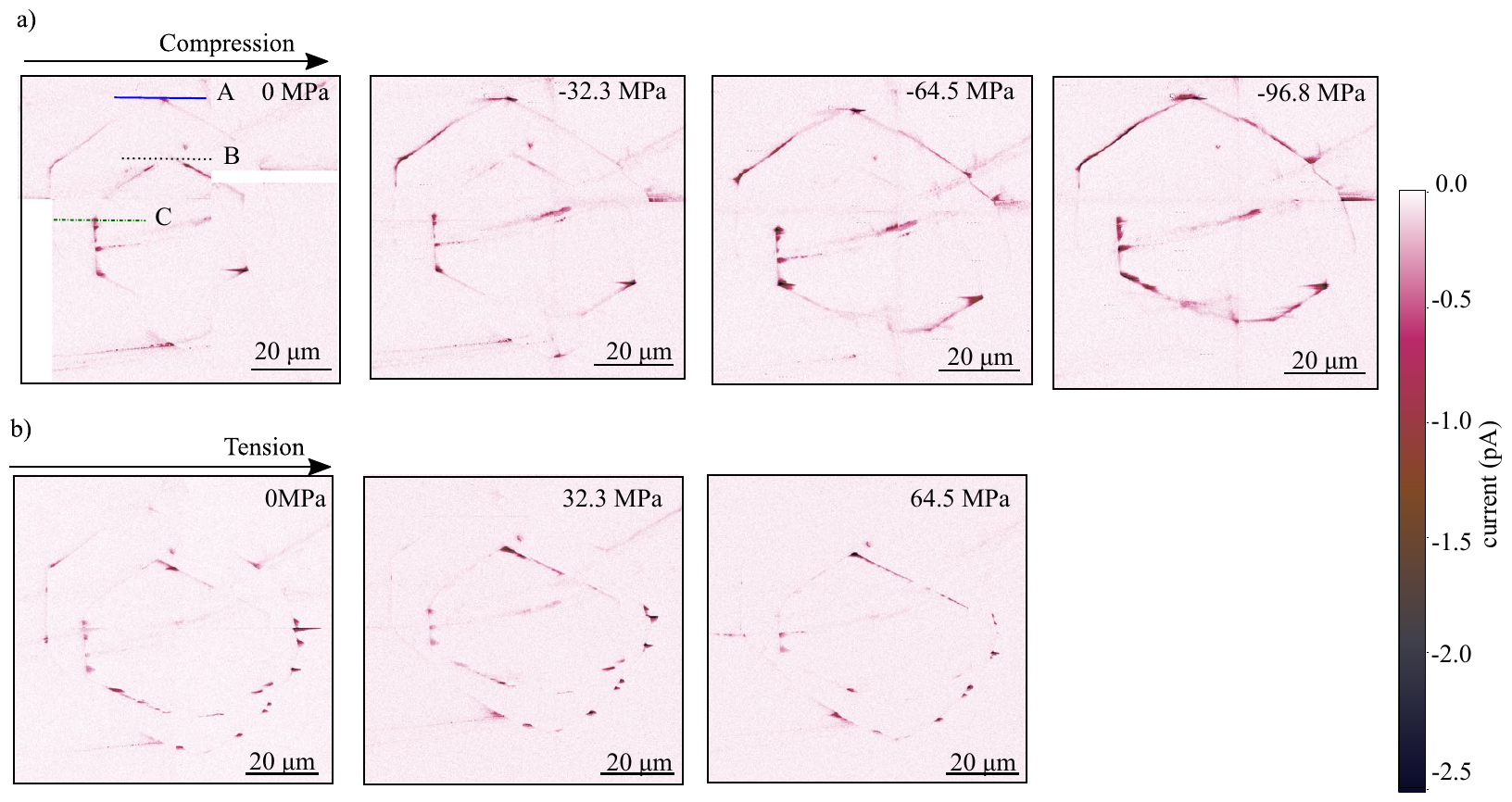}
\caption{\label{fig:micro_result}cAFM scans of sample LNO-01 for compression from 0 MPa to -96.8 MPa and for tension from 0 MPa to 64.5 MPa, showing how current is evolving in different parts of domain walls on compression and tension.}
\end{figure*}

In Fig.~\ref{fig:micro_result} we have provided all the scans of sample LNO-01 between 0 MPa to -129 MPa for compression and from 0 MPa to 64.5 MPa for tension.

\begin{figure}[h!]
\includegraphics{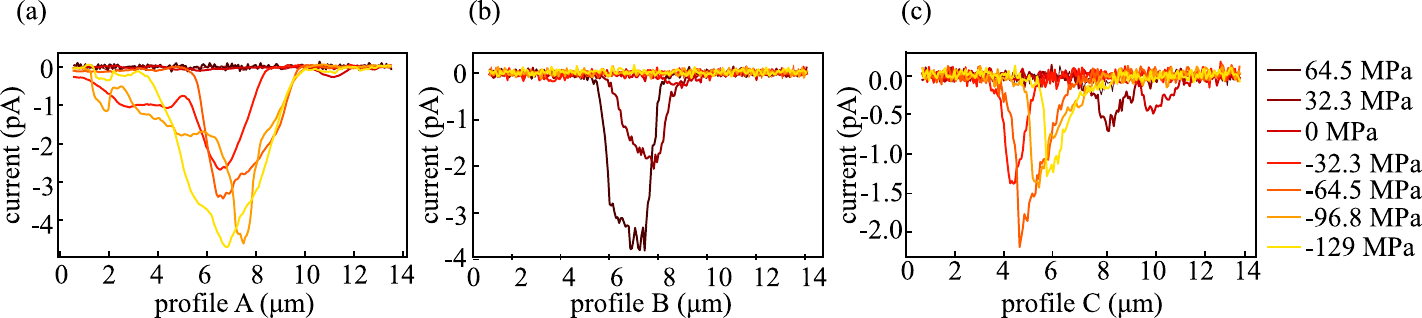}
\caption{\label{fig:micro_profile_result} Current profiles of line profile A,B, and C from different sections of domain walls in Fig.~\ref{fig:micro_result} for sample LNO-01.}
\end{figure}

In Fig.~\ref{fig:micro_profile_result} we show the profiles of the line A, B, C from different sections of the domain walls provided in Fig.~\ref{fig:micro_result}. Subfigure (a) depicts increase of current with compression, subfigure (b) shows decrease in current with compression and subfigure (c) shows no clear trend. We have taken the maximum of all these curves and have plotted them as a function of stress at x-axis of graph in Fig. 4 of the main paper.

\begin{figure}[h!]
\includegraphics{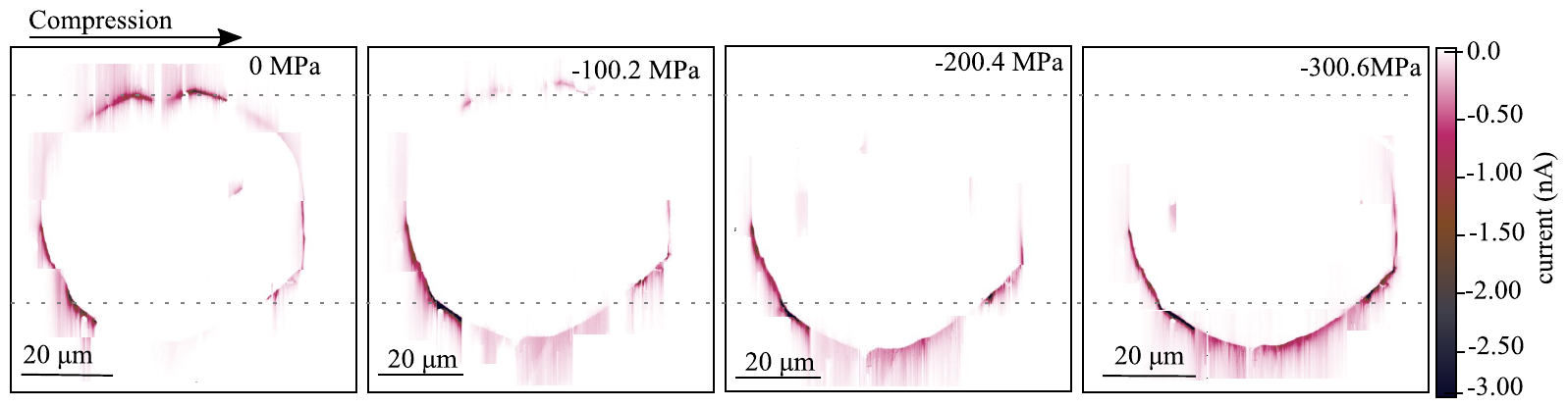}
\caption{\label{fig:micro_result_1}cAFM scans of sample LNO-02 for compression from 0 MPa to -300.6 MPa, showing how current is evolving in different parts of domain walls on compression.}
\end{figure}

In Fig.~\ref{fig:micro_result_1} we show compression trends of sample LNO-02 from 0 MPa to -300.6 MPa in the steps of 100.2 MPa.

\subsection{S3.2: stability and repeatability measurements}
\begin{figure*}[!h]
\includegraphics{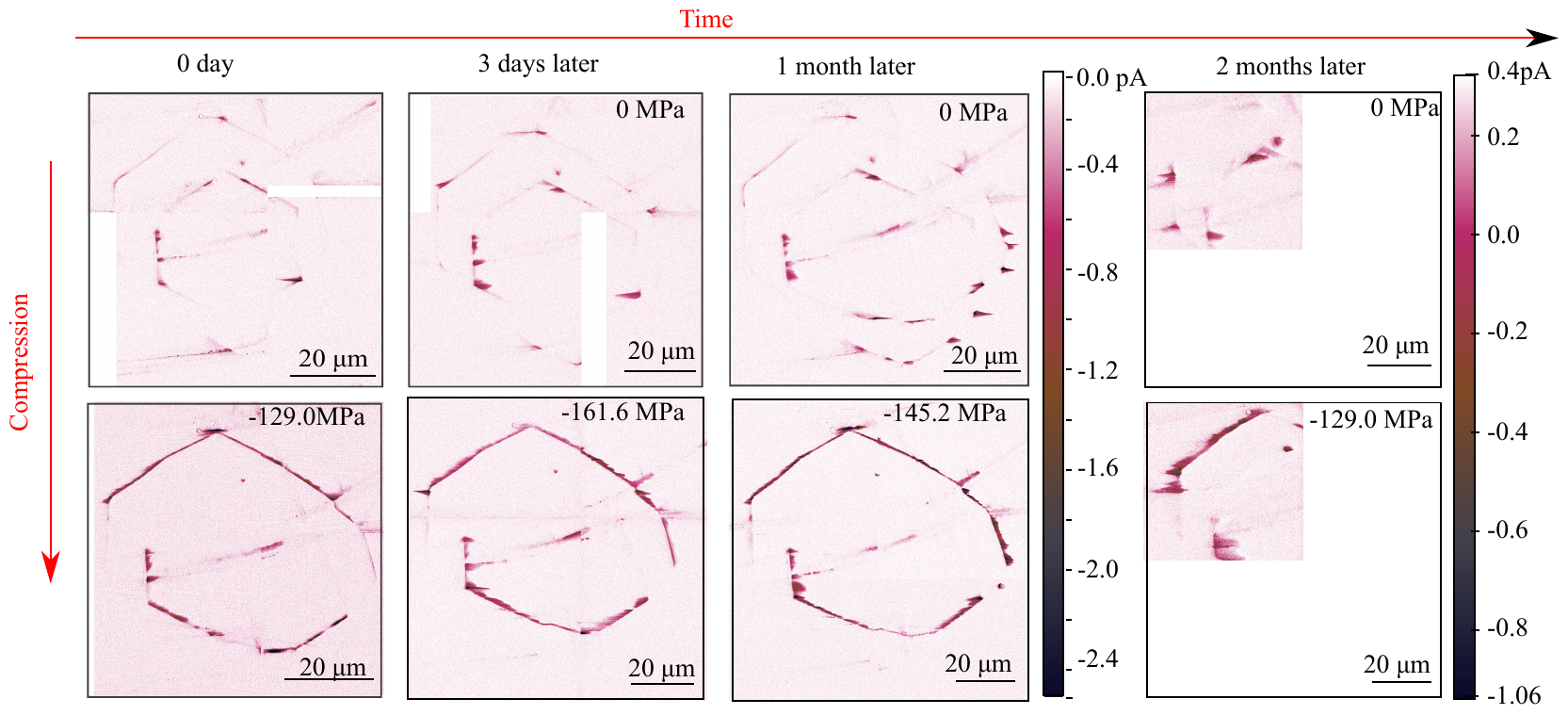}
\caption{\label{fig:stability} cAFM scans of sample LNO-01 over the time of 2 months at 0 MPa and at different compressive stress values. The results show that the measurement is repeatable.}
\end{figure*}

In Fig.~\ref{fig:stability} we show based on the sample LNO-01 that the measurement is stable and repeatable over the time of 2 months. Although, we see a overall decrease in the current after two months atleast the qualitative behaviour with the stress remains intact.

\subsection{S3.3: Measurements performed at +10 V cAFM-scans}
\begin{figure}[h!]
\includegraphics{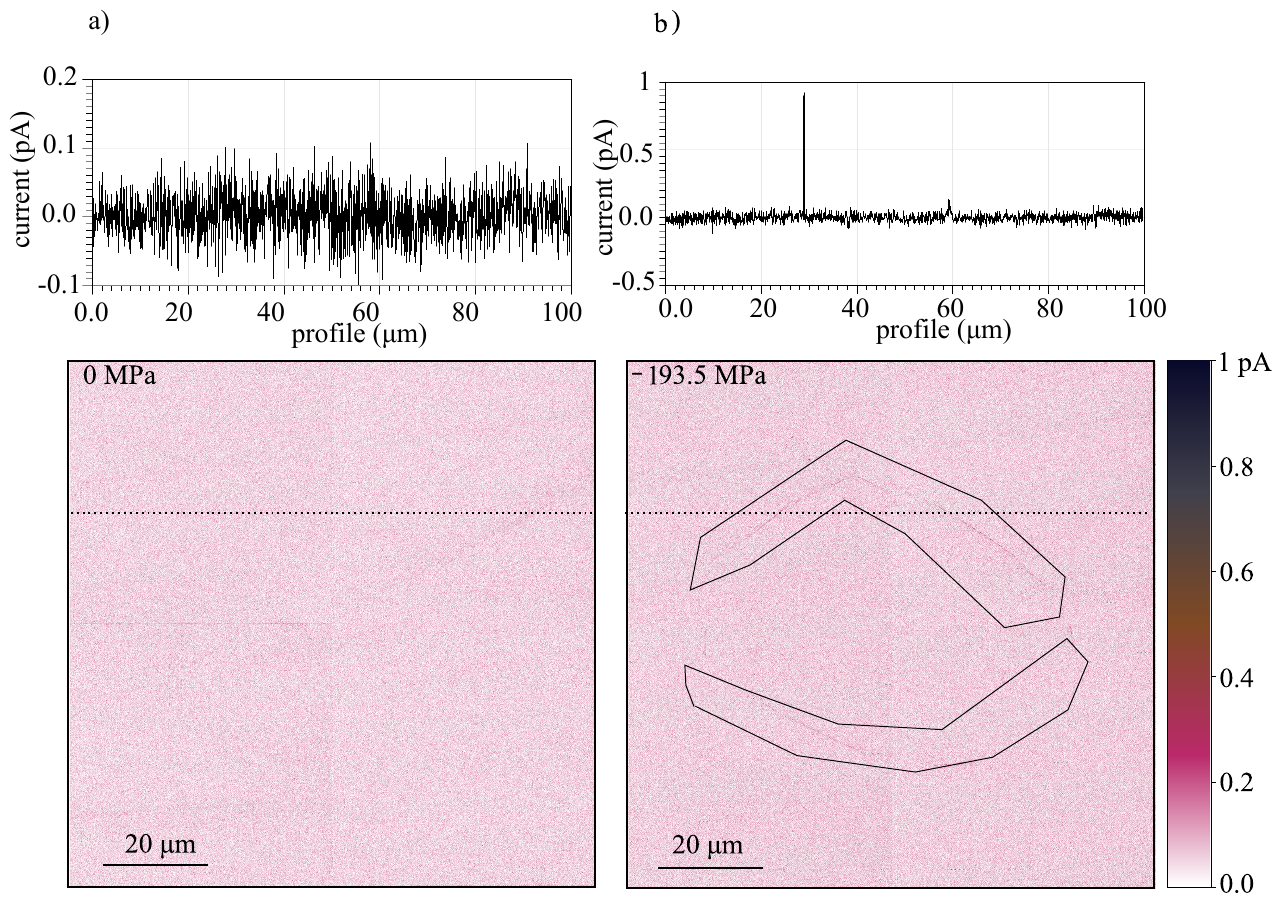}
\caption{\label{fig:10V_scan}(a) cAFM scan of unstressed LNO-01 sample at +10 V shows no current at domain wall, confirmed by the profile data above that there is no distinctive peak for domain wall current (b) cAFM scan on sample compressed, to -193.5 MPa, at +10 V shows very small current at domain wall, confirmed by the profile data above the scan. The irregular curves are shown to point the region of interest.}
\end{figure}
As it can be seen in I-V curves of sample LNO-01 in Fig.~\ref{fig:s1} (a), the sample shows current in the order of pA at +10V, that is why the cAFM experiment was performed at -10 V. Nevertheless when the same experiments were performed at +10 V, no change in current was observed until $\approx$-100 MPa. For larger stress values very faint lines of current start getting visible as can be seen in Fig.~\ref{fig:10V_scan}(inside the irregular boxes) and also through profiles on top of the images. Subfigure (a) shows there is no current within the limit of the amplifier at 0 MPa while, subfigure (b) shows that a masureable current starts to appear at  the upper part of the outer domain and lower part of the inner domain, as was seen at -10 V.